\documentclass[aps,twocolumn,superscriptaddress]{revtex4-1}
\pdfoutput=1
\usepackage{graphicx}
\usepackage{epstopdf}
\usepackage{amssymb}
\usepackage{mathrsfs}
\usepackage{amsmath}
\usepackage{color}
\usepackage{latexsym}
\usepackage{amsfonts}
\usepackage{hyperref}
\usepackage{subfigure}
\usepackage{wasysym}
\usepackage{dcolumn}
\usepackage{bm}
\usepackage[percent]{overpic}
\usepackage{natbib}

\begin{document}

\selectfont
%\title{Evidences for reentrant liquid-to-gas Spinodal in colloidal systems}
\title{Reentrant spinodals in colloidal model systems}
\title{Reentrant spinodals and the Speedy scenario in colloidal model systems}

\author{Lorenzo Rovigatti}
\affiliation{Rudolf Peierls Centre for Theoretical Physics, 1 Keble Road, Oxford, OX1 3NP, UK}

\author{Valentino Bianco}
\affiliation{Faculty of Physics, University of Vienna, Boltzmanngasse 5, A-1090 Vienna, Austria}

\author{Jos\'e Maria Tavares}
\affiliation{Instituto Superior de Engenharia de Lisboa - ISEL,
Rua Conselheiro Em\'{\i}dio Navarro 1, P-1950-062 Lisbon, Portugal }
\affiliation{
Centro de F\'{\i}sica Te\'{o}rica e Computacional, Universidade de Lisboa,
Campo Grande, P-1749-016 Lisbon, Portugal
}

\author{Francesco Sciortino}
\affiliation{Dipartimento di Fisica, Sapienza-Universit\'a di Roma, Piazzale A. Moro 5, 00185 Roma, Italy}
\affiliation{Istituto Sistemi Complessi (CNR-ISC), Via dei Taurini 19, 00185 Roma, Italy}

\definecolor{corr14okt}{rgb}{0,0,1} % blue
\definecolor{moved}{rgb}{0,0,0}

\begin{abstract} 
A re-entrant gas-liquid spinodal was proposed  
as a possible explanation of the apparent divergence of the compressibility and specific heat
on supercooling water.  Such a counter intuitive possibility, \textit{e.g.}  a liquid that becomes unstable to gas-like
fluctuations  on cooling at positive pressure, has never been observed, neither in real substances nor 
in off-lattice simulations.  More recently, such re-entrant scenario   has been  dismissed 
on the premise that  the re-entrant spinodal  would collide with the gas-liquid binodal in the pressure-temperature plane.    Here we study, numerically and analytically,  two
previously introduced  one-component patchy particle models that both show (i) a  re-entrant  spinodal and (ii) a re-entrant  binodal, providing a neat {\it in silico} (and {\it in charta})  realization of such
unconventional  thermodynamic scenario. 
 \end{abstract} 

\maketitle

Liquids with isobaric density extrema are particularly fascinating. The tendency to expand on cooling
signals the onset of a structural ordering at microscopic level that overrides the ubiquitous densification originating from the
  reduced amplitude of thermal vibrations.  The investigation of the thermodynamic behavior of
  liquids with density anomalies has clarified that  density extrema are never isolated
  anomalies, but are always associated with non-monotonic behaviors of several other response functions.  In the case of water --- the most common member of the group, whose  temperature ($T$) of maximum density (TMD)  at ambient pressure ($P$) is $4^\circ$C --- experimental   studies have shown that,  upon cooling, density, energy and entropy fluctuations all increase, leading to an anomalous growth of  the constant-$P$   compressibility and specific heat~\cite{kanno:4008, AngellC.A._j100395a032, debenedetti-book}. Such growth is further enhanced in supercoled states, where water is metastable with respect to crystallization.

\begin{figure}
\includegraphics[scale=1]{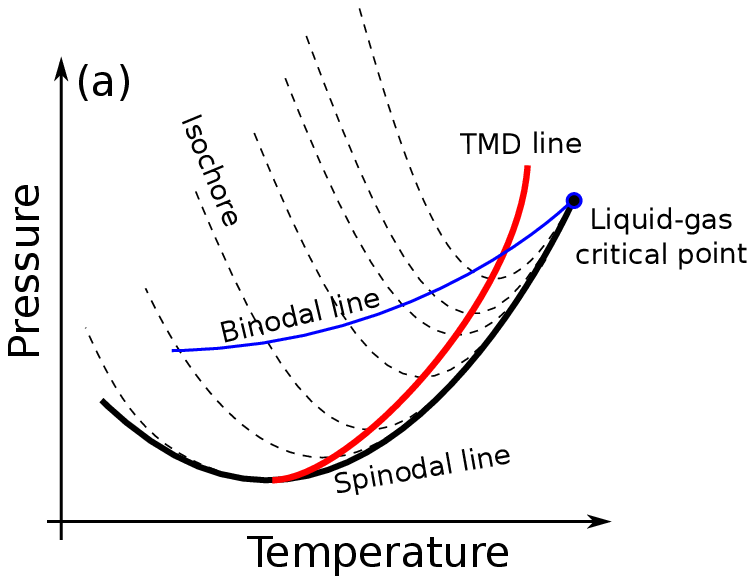}
\includegraphics[scale=1]{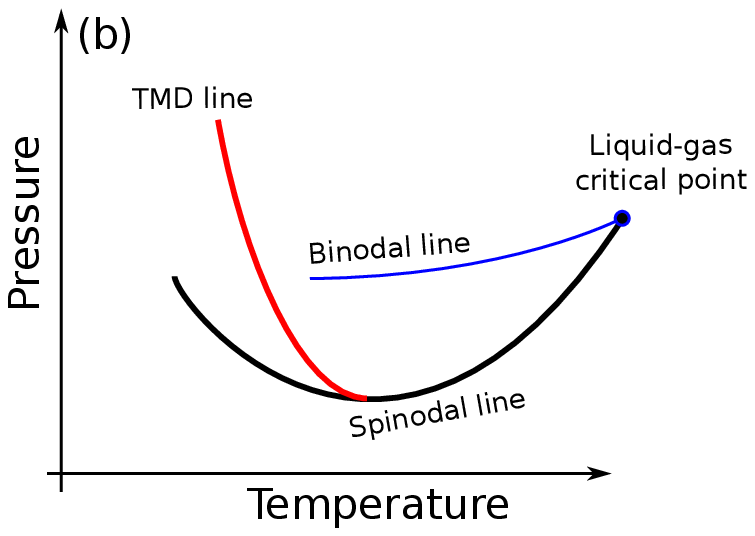}
\caption{(a) The reentrant liquid-gas spinodal scenario. The TMD line connects all the minima of the isochores where $(\partial P/\partial T)_V=0$ and intersects the spinodal line in its turning point.
(b) The reentrant liquid-gas spinodal scenario with negative sloped TMD line. The slope of the TMD line is related to the isobaric behavior of the isothermal compressibility $K_T=-(1/V)(\partial V/\partial P)_T$~\cite{Sastry1996}, in turn proportional to the density fluctuations. If the TMD has a negative/positive slope in the $T$--$P$ plane, then $K_T$ increases upon cooling/heating. % In water the TMD line has a negative slope when $P>0$ and a negative slope when $P<0$.
}
\label{fig:spinodal_scheme}
\end{figure}

The origin of density anomalies in water has animated the debate in the scientific 
community~\cite{kanno:4008, AngellC.A._j100395a032, poolePRL1994, Tanaka96, Mishima1998, SoperPRL2000,  Xu2005, LiuPNAS2007, StokelyPNAS2010, abascalJCP2010, bertrand2011, holtenJCP2013, azouzi2013coherent, pallares2014anomalies, BiancoSR2014, Soper2014, Nilsson2015}. Different thermodynamic consistent scenarios have been proposed~\cite{Speedy1982, 
Sastry1996, llcp, AngellScience2008} leading to intense discussions  which extends up to present days~\cite{sciortinoPCCP2011, liuJCP2012, palmer_car_debenedetti, pooleJCP2013, Palmer2014, SmallenburgPRL2015}.
The first  thermodynamic scenario coherently accounting for the observed  density, compressibility and
specific heat anomalies of water was proposed in 1982 by Robin Speedy~\cite{Speedy82}.  In this very elegant piece of work,
Speedy focused on the limit of stability of the liquid phase --- which in mean-field coincides with the
gas-liquid spinodal line --- a line emanating from the gas-liquid critical point. In standard liquids, as predicted by the
venerable van der Waals theory,  the gas-liquid spinodal is monotonic in the $P-T$ plane
approaching $T=0$ at negative $P$, at the maximum tensile strength value~\cite{debenedetti-book}. Speedy noticed
that the spinodal line $P_{spinodal}(T)$ is an envelope of isochores~\cite{debenedetti-book, poolePRE1993}  and thus a TMD locus which intersects the spinodal line  requires $dP_{spinodal}/dT=0$  at the intersection, e.g.
 a re-entrant behavior.  Fig.~\ref{fig:spinodal_scheme} shows a sketch of the Speedy phase diagram for positively and negatively sloped TMD loci.  Albeit highly counterintuitive,  if the spinodal traces back to positive $P$, 
the liquid becomes unstable to gas-like fluctuations both on heating and on cooling.  In Speedy's scenario, the encounter of the
retraced spinodal  on cooling explained the observed increase in the response functions. 
   
In 2003,  Debenedetti~\cite{De03} called attention on the fact that a  reentrant  spinodal can not intersect  
 the metastable continuation of the liquid-gas binodal line without terminating there in an additional critical point.
In the case of water, this would suggest that the response function should not display any significant increase for pressure 
sufficiently higher than the triple point pressure, at odd with experimental observation. 
The interest in finding realizations of the Speedy reentrant spinodal has been progressively attenuated, if not suppressed, by the difficulty to imagine a liquid that would vaporize on cooling and by the availability of different thermodynamic scenarios equally able to rationalize the anomalies (the liquid-liquid critical point~\cite{llcp}  and the singularity-free~\cite{Sastry1996}  scenarios).   With the exception of lattice models of water-like fluids~\cite{Sastry1993, Sasai1993, Borick1995}, where the spinodal was found to retrace but only at negative pressures,  Speedy's hypothesis thus remains a fascinating scenario that has never been realized, not even {\it in silico}.

In this Letter we  fill this gap, reporting  two examples of a one-component system  exhibiting reentrant spinodal, 
covering both the cases presented in Fig.~\ref{fig:spinodal_scheme}, e.g. with 
positively and negatively sloped TMD respectively.
Interestingly, in both cases,  the intersection between the re-entrant spinodal and the gas-liquid binodal is avoided, providing support to the Speedy's scenario and its compatibility Debenedetti's arguments.

The systems considered here are composed of spherical  hard particles complemented with anisotropic attractions.
  In both cases, particles are modelled as hard spheres of diameter $\sigma$ (the unit of length). Each particle is decorated with $n_p$ patches, which are modelled as truncated spherical cones and interact between themselves through a Kern-Frenkel (KF) potential~\cite{Kern2003}, an angular square well of depth $\epsilon$ (the unit of energy).  
Both models are examples of \textit{patchy}  particles~\cite{Bianchi2011}, colloids of new generation that  
have been shown to exhibit interesting and unique states such as empty liquids~\cite{Bianchi2006}, open crystals~\cite{flavio_tetra,doppelbauer_patchy_genetic}, water-like liquid-liquid critical points~\cite{smallenburg2014erasing} and more~\cite{Wilber2009,delasheras_bicontinuous_gels,Coluzza2013}.
\begin{figure}[h]
%\centerline{
%\includegraphics[scale=1.2]{t-rho_scheme_2A9B.eps}
\begin{overpic}[scale=1]{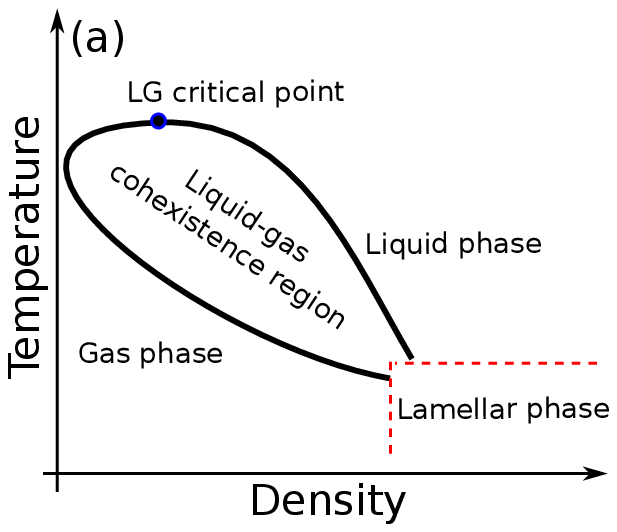}
 \put(75,55){\includegraphics[height=2cm]{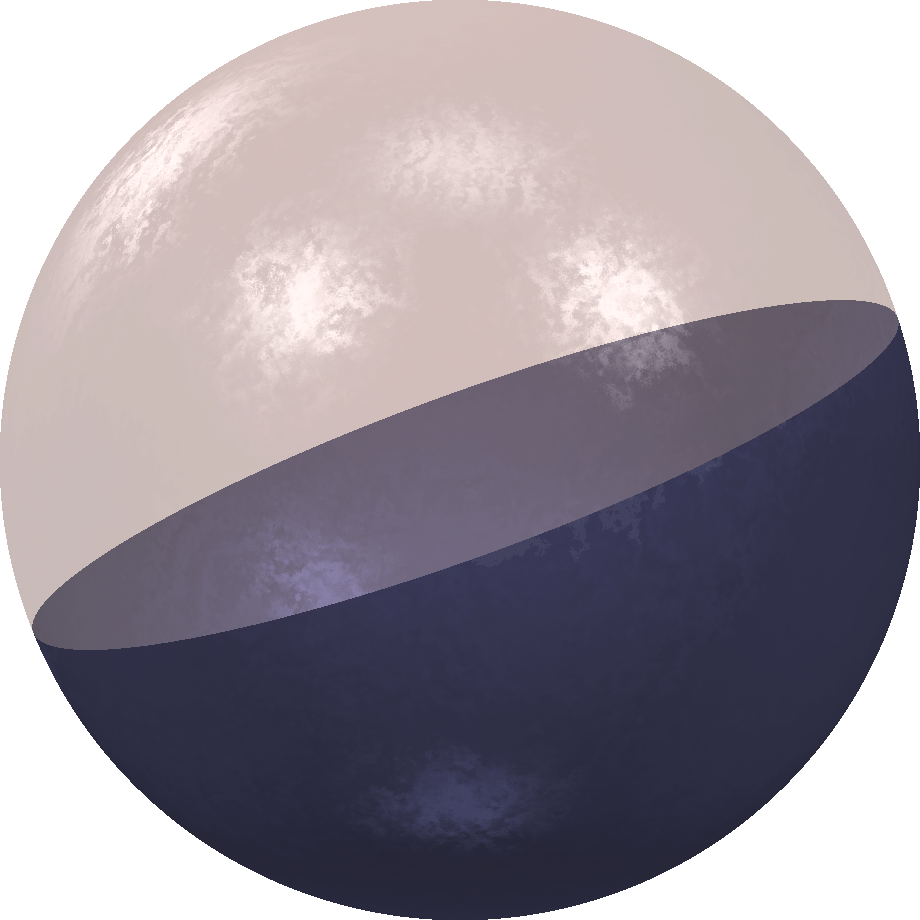}}
\end{overpic}\\
\vspace{0.5 cm}
\begin{overpic}[scale=1]{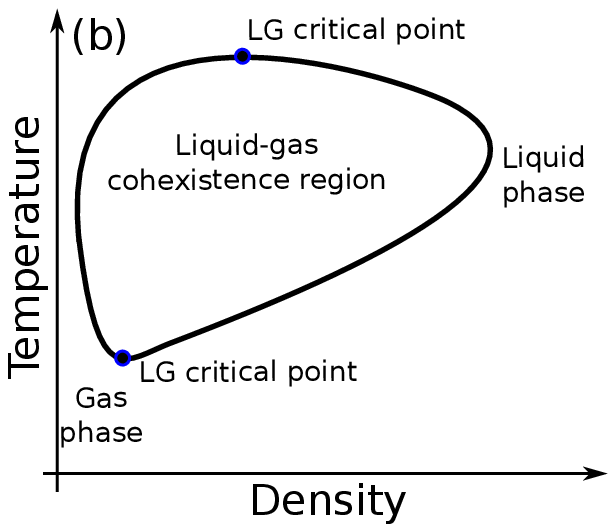}
 \put(75,15){\includegraphics[height=2cm]{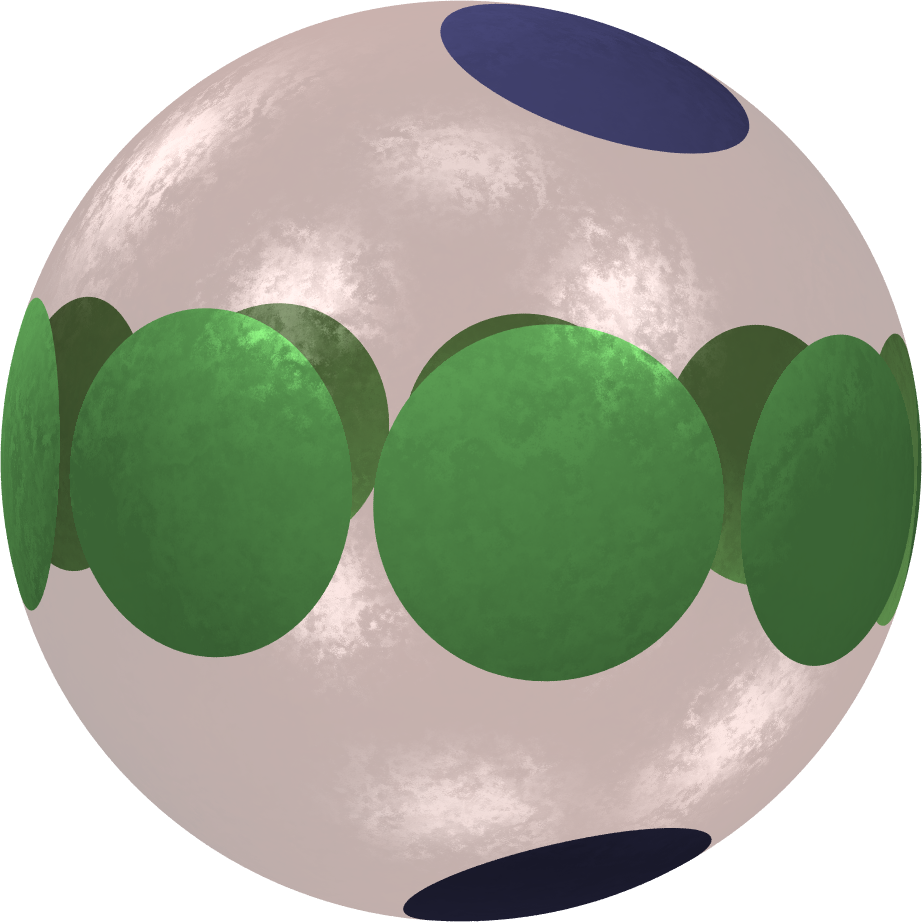}}
\end{overpic}
\caption{(a) Sketch of the phase diagram in the $\rho$--$T$ plane for the Janus system, composed by particles that are hard spheres with an attractive patch covering half of its surface (in blue). (b) Sketch of the phase diagram of the 2A9B system. The surface of these particles is decorated with two patches of type A (in blue) located close to the poles, so that the relative patch-center-patch angle is $130^\circ$, and nine patches of type B (in green) equispaced on the equator.}
\label{fig:models}
\end{figure}

Figure~\ref{fig:models} shows a cartoon of the two models and the associated schematic phase diagram.  Both models
have been previously introduced and their phase diagram in the $T$-$\rho$ plane evaluated. Here we return to these two
models evaluating the equation of state and the TMD and spinodal loci in the $P-T$ plane (see Supplemental Information, SI, for a detailed presentation of the two models).
The first model we consider is a Janus particle 
composed by two different hemispheres, one repulsive and one attractive. The hard sphere is decorated  with a single patch covering half of its surface.
The phase behaviour of Janus colloids is extremely rich, featuring non-crystalline ordered phases, cluster phases, crystalline lattices and a  gas-liquid phase transition~\cite{Hong2008,janus_pd,janus_crystals,janus_sedimentation}. The latter has a peculiar shape as the two coexisting densities, gas and liquid alike, increase upon cooling. However, the coexisting gas density increases faster, leading to the shrinkage, and possibly closure, of the unstable region. However, the seemingly inevitable appearance of a lower critical point is prevented by the presence of an ordered lamellar phase (see Fig.~\ref{fig:models})~\cite{janus_pd}.
In the second model~\cite{russo2011reentrant,closed_loop}, indicated as 2A9B  in the following, the  hard sphere is decorated with two patches of type A, located on opposite hemispheres in such a way that the patch-particle center-patch angle is $\gamma$, and nine patches of type B, equi-spaced on the equator. Bonds between B patches are disabled. The attraction strengths of the   interactions are chosen in such a way that the formation of AA bonds, and hence of long chain-like structure, is energetically favoured. However, the larger number of B-patches makes AB bonds entropically convenient, promoting the branching of the chains. When $\gamma = 180^\circ$ the competition between the two mechanisms results in a pinched gas-liquid phase diagram, where the density of the liquid phase decreases as temperature goes down~\cite{russo2011reentrant}. For smaller values of $\gamma$, however, the chains are much more flexible and the system tends to form rings at very low $T$. These rings stabilise the gas phase, causing a reentrance of the density of the gas which approaches the coexisting liquid one. The net effect is to generate a closed gas-liquid coexistence loop in the $T-\rho$ plane with two  critical points (see Fig.~\ref{fig:models})~\cite{closed_loop}. Here we choose $\gamma = 130^\circ$. 
The Janus and 2A9B models investigated here display gas-liquid phase separation regions where, at low $T$, the $\rho$ of one or both of the two phases does not tend to a constant value, but rather continuously increases or decreases as shown in Fig.~\ref{fig:models}. As a result, the two phases approach each other, and the phase diagram displays a \textit{reentrant} shape. As we will show in the following, such a phenomenon, which is linked to the peculiar self-assembly processes occurring in these systems, is also connected to thermodynamic anomalies. 

To evaluate  $P(\rho,T)$, we perform simulations in the grand~canonical ensemble with the successive umbrella sampling (SUS) technique~\cite{sus}, which consists in an independent sampling of overlapping density intervals. The results obtained in each interval at fixed  $T$, volume $V$ and chemical potential $\mu$ are then stitched together by using the overlapping regions between neighbouring windows~\cite{sus}. 
The main simulation output is the probability that the system has a number density $\rho = N/V$ (with $N$ the number of particles in the system), namely ${\cal N}(\rho)$, computed at fixed $V$, $T$ and $\mu$ (or, equivalently, fixed activity $z = e^{\beta \mu}$, where $\beta = 1/k_BT$ and $k_B$ is the Boltzmann constant). 
The uniform (in $N$) sampling provided by the SUS technique allows us to evaluate ${\cal N}(\rho)$ at different $\mu$ by employing histogram reweighting techniques\cite{histogram_reweight}.

\begin{figure}%[h!]
\includegraphics[width=0.45\textwidth]{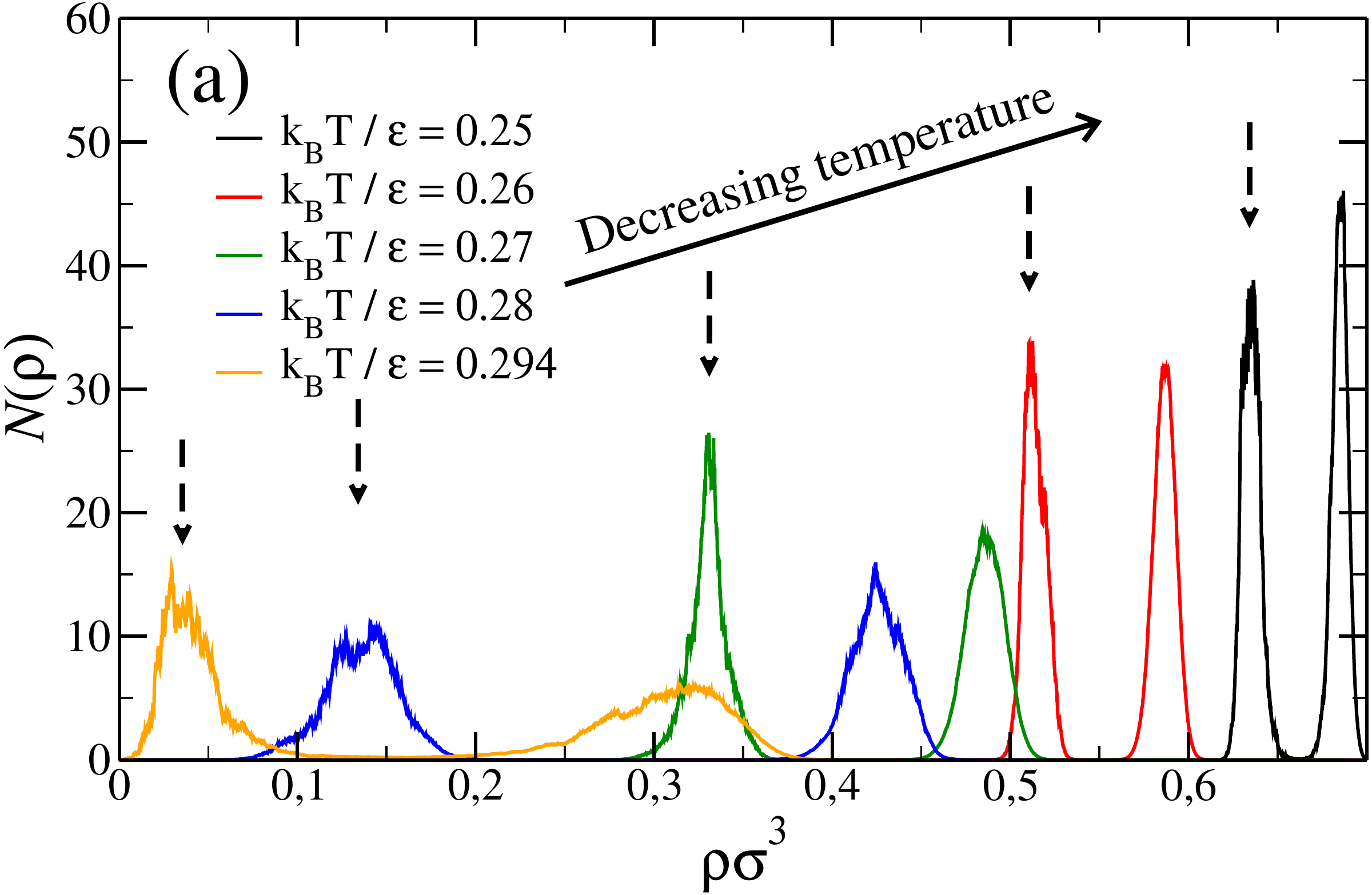}\\
\includegraphics[width=0.48\textwidth]{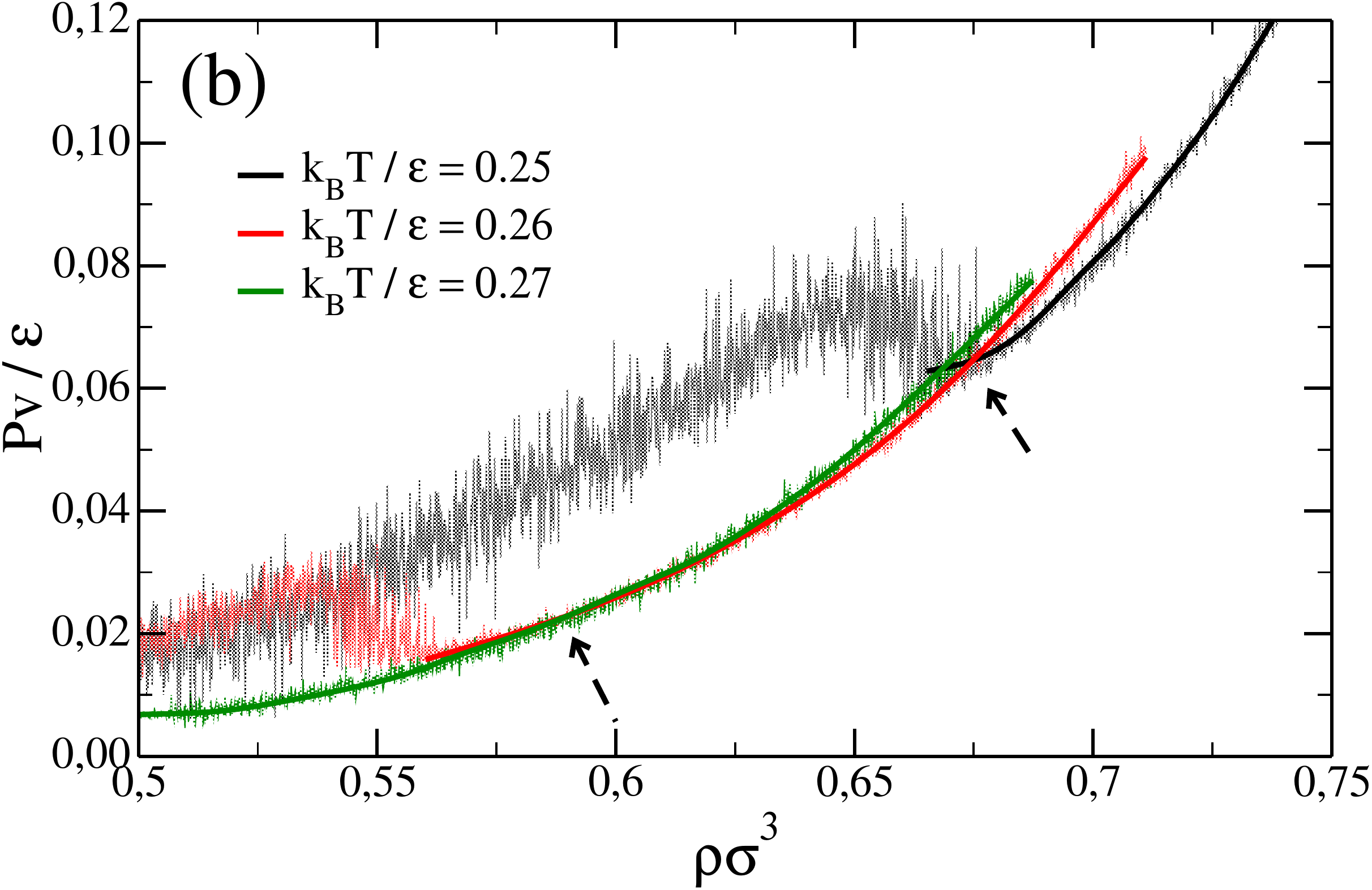}
\caption{\label{fig:prho}(a)
Probability distributions of the density, ${\cal N}(\rho)$, for the Janus system at five different $T$. The bimodality of the curves demonstrates that, at all these $T$, there is a phase separation between a low-density (gas) and a high-density (liquid) phase. The vertical dashed arrows indicate the position of the peak relative to the gas phase, which moves to higher and higher density as $T$ decreases. The value of the chemical potential at each $T$ has been chosen so that the area below each curve is equally shared between the two peaks, which is the condition of phase coexistence.  The simulation box has a side $L=20 \sigma$, requiring the investigation of 
a number of particles extending from 0 up to 5000.
(b) Equations of state $P(\rho,T)$ for the Janus system at three different $T$. The raw data (shaded curves) is splined under tension. The intersections between the resulting curves (solid lines), here indicated with arrows, yield estimates for the location of the density maxima.}
\end{figure}

As an example, Figure~\ref{fig:prho}(a) shows  the ${\cal N}(\rho)$ curves obtained for the Janus system. All the curves have been reweighted at coexistence, which is reached when the area below each of the two peaks is the same.
From ${\cal N}(\rho)$, $P(\rho,T)$ can be directly calculated (see SI). As shown by Binder and his group~\cite{binder_spinodal}, when simulations are properly performed and the system has been able to equilibrate even inside the coexistence region, $P(\rho,T)$ provides a consistent  thermodynamic description valid for the  investigated system size. From the resulting $P(\rho,T)$ we thus estimate the coexistence pressure as well as the lowest pressure on the liquid branch that we define as a proxy of the
mean-field spinodal. Finally, from the intersection of the curves $P(\rho,T)$ for different values of $T$ 
we determine the presence of isobaric extrema of the density $\rho$ (Fig. \ref{fig:prho}b, SI).

\begin{figure}%[h!]
\includegraphics[width=0.47\textwidth]{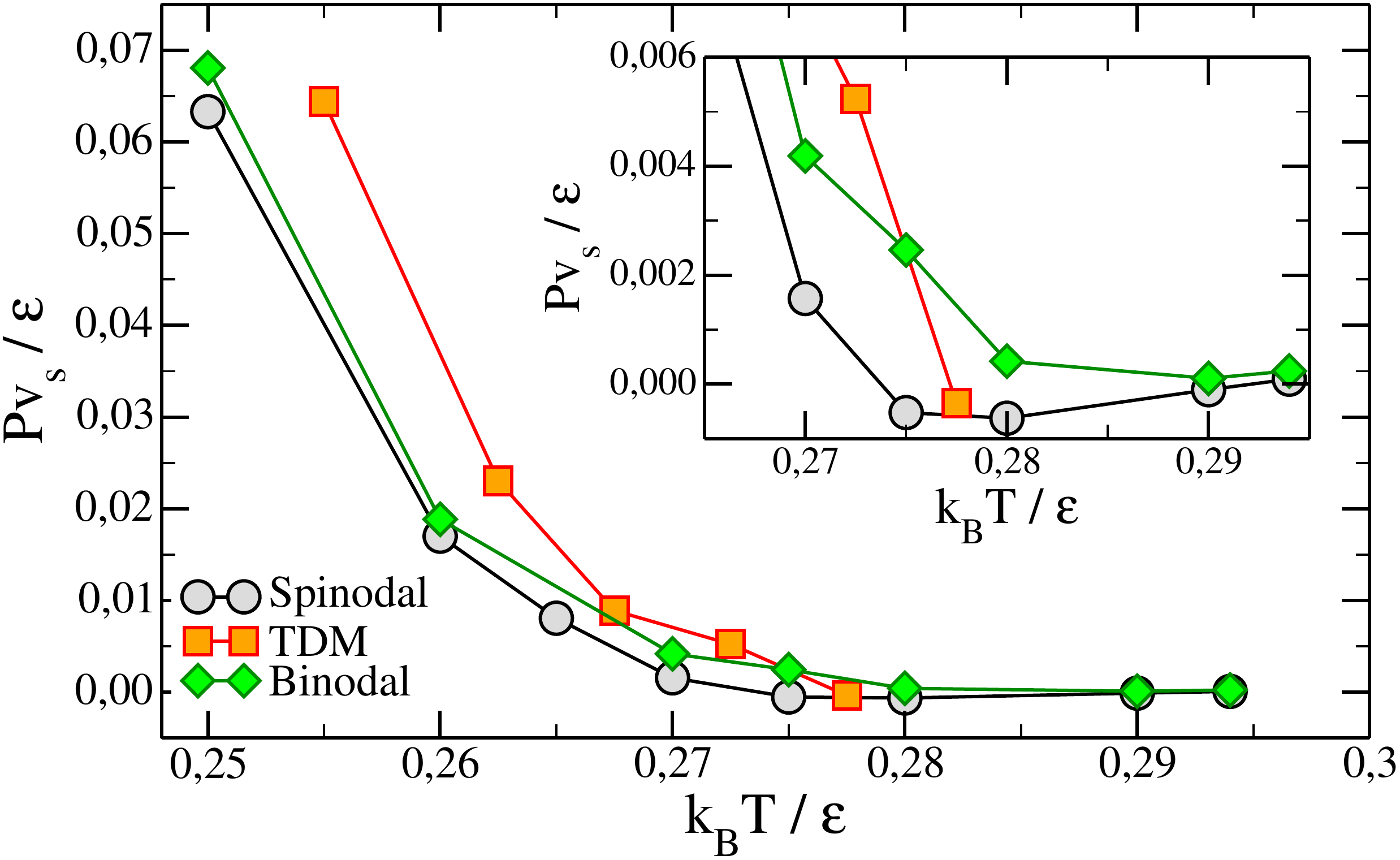}
\caption{\label{fig:PT_janus}The spinodal (gray circles), TMD (orange squares) and binodal (green diamonds) loci of the Janus model in the $P-T$ plane. (Inset) A blow-up of the main plot, showing that the TMD line ends in the minimum of the spinodal. Lines are guides for the eye.}
\end{figure}

Figure~\ref{fig:PT_janus} shows the binodal, spinodal and TMD lines, as evaluated for the Janus model in the $P-T$ plane. We first note that, at high $T$, the spinodal and binodal lines originate from  the critical point. As $T$ decreases,  the binodal $P$ remains roughly constant, while $P_{spinodal}$ decreases, becoming negative. At around $k_BT/\epsilon  \simeq 0.2775$ a minimum in $P_{spinodal}$ appears, and then both the spinodal and binodal $P$ quickly raise up as the system is further cooled down. 
By comparing these results with the $T-\rho$ phase diagram~\cite{janus_pd} we link the observed sudden increase of the coexistence pressure on cooling to the increasing of the coexisting density of the gas. Indeed, in this $T$ region the coexisting gas, which at higher $T$ is a diluted phase made of mostly non-interacting particles, turns into a cluster phase whose basic constituents, micelles and vesicles, are finite-sized aggregates of particles~\cite{janus_pd}. The return of the spinodal line to the $P>0$ semi-plane, which happens around $k_BT/\epsilon\approx 0.274$, is the first numerical example of a 3D off-lattice system exhibiting a Speedy-like scenario.

Our results show that the Janus model also exhibits a line of density maxima, as required by thermodynamic consistency~\cite{poolePRE1993}. This TMD line, shown in Figure~\ref{fig:PT_janus} has a negative slope, as in water~\cite{henderson1987temperature}. It starts from high pressure at low temperature and quickly goes down, intercepting the spinodal line. The two lines meet, within our numerical accuracy, exactly at the minimum of the spinodal, as highlighted in the inset of Figure~\ref{fig:PT_janus}. 
At the intersection point, the TMD line shows a vanish slope, consistently with thermodynamic predictions~\cite{poolePRE1993}.
It is interesting to note that the sheer existence of a binodal with a negative slope avoids the argument that a reentrant spinodal must intersect the binodal line if it retraces its path to positive pressures~\cite{De03}.

\begin{figure}%[t!]
\includegraphics[width=0.47\textwidth]{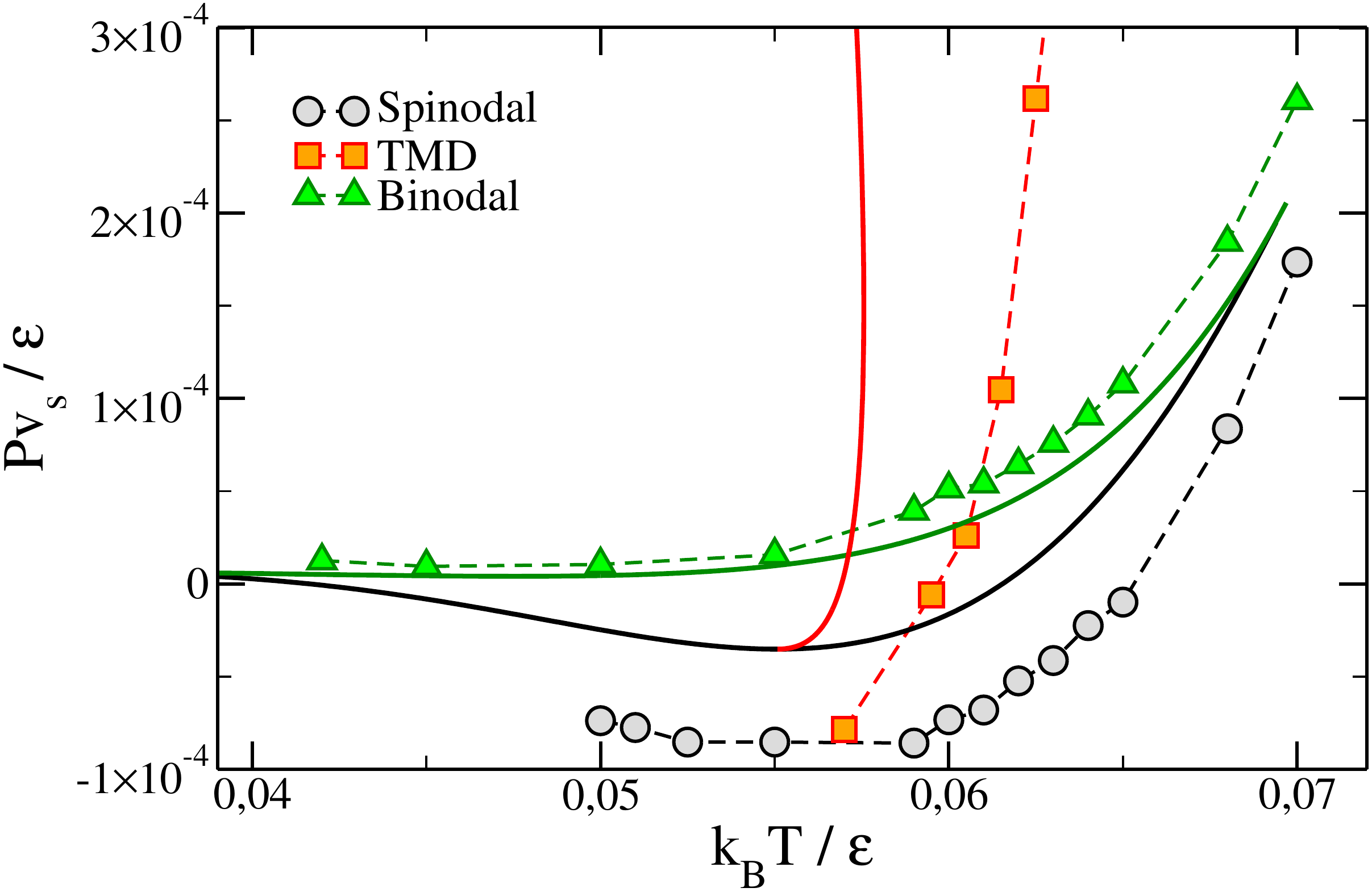}
\caption{\label{fig:PT_2A9B}Theoretical and numerical results for the 2A9B model:
 spinodal (black), TMD (orange), and binodal (green) loci in the $P-T$ plane. 
 Symbols indicate numerical results, while theoretical curves are shown as full lines. Dashed lines are guides for the eye.
 Numerical results refer to a system with $L=14\sigma$.}
\end{figure}

Next we focus on the 2A9B model.
%, whose original implementation ($\gamma = 180^\circ$) was shown to exhibit a reentrance of the density of the coexisting liquid in the $T-\rho $ plane~\cite{russo2011reentrant}.
 To favour the formation of energetically stable weakly interacting aggregates
 (rings as opposed to chains, e.g.  structures in which all AA bonds are satisfied)
  in the gas-phase  we select $\gamma = 130^\circ$.  This provides thermodynamic
stability to the gas phase~\cite{closed_loop}. Figure~\ref{fig:PT_2A9B} shows the thermodynamics \textit{loci} of the model in the $P-T$ plane. In contrast to the Janus model, the pressure of the binodal (for $k_BT/\epsilon > 0.045$) and of the TMD is an increasing (rather than decreasing) function of $T$. 
Similarly to the Janus case, the liquid spinodal is also non-monotonic, with a minimum occurring at $k_BT/\epsilon \approx 0.055$. As a result, the TMD and the spinodal curves meet, within our numerical accuracy, in the minimum of the latter, consistently with thermodynamic predictions. In contrast with the Janus case, however, the TMD line has a positive slope.

The 2A9B model with $\gamma = 130^\circ$ can also be solved in mean-field~\cite{closed_loop}, within the Wertheim formalism~\cite{Werth1}, augmented with the inclusion of closed ring loops~\cite{tavares,closed_loop}.  The analytic solution provides an unambiguous definition of the spinodal line and thus implicitely a check of the numerical definition.
 The theoretical results, shown as full lines in the figure, compare well with simulation data in the $T$-range where the latter is available. At lower $T$ the theory shows that the spinodal curve returns to positive $P$,  fully  consistent with  Speedy's scenario. According to theory (and with Debenedetti's arguments), the spinodal line ends at the lower critical point ($T\approx 0.0339$), where it meets with the binodal.

To summarise, a retracing LG spinodal was proposed for the first time almost 35 years ago by Speedy to explain the low-temperature anomalies of water~\cite{Speedy1982}. In this Letter, for the first time, we have presented two off-lattice examples of systems exhibiting such a unique thermodynamic feature.  In both cases, the liquid vaporizes on cooling, but
the gas phase is composed by (weakly interacting) aggregates in which particles are ordered in configurations of
very low energy but also very low entropy,  changing the slope of the binodal at low $T$. In the Janus case, the gas is indeed formed by micelles and vesicles~\cite{janus_pd}, while
in the 2A9B case, the gas is composed by rings~\cite{closed_loop}. Thus, the physics that stabilizes the gas phase at low $T$ allows the spinodal to rise back to positive $P$. 
In the case of water, in the region where experiments are possible, there is no evidence of a lower gas-liquid critical point and the binodal is positively-sloped, ruling out 
the Speedy scenario. 
Finally, we note that both anisotropically interacting patchy particles models can possible be realized experimentally in the near future~\cite{Zhang2009,Andala2012,Wang2012e,sacanna_review},  allowing for an experimental confirmation of the numerical and theoretical results reported here. The peculiar properties of these systems, while being somewhat different from those of water, can shed light on the thermodynamics of anomalous fluids. We have found that the observed non-monotonic behaviour of the LG spinodal line is linked to the reentrance of the density of the coexisting gas (Fig. \ref{fig:models}). This strongly suggests that such a reentrance is a sufficient condition to observe a non-monotonic spinodal line and, as a consequence of thermodynamic consisentecy, a TMD line which meets the spinodal exactly in its minimum.

\section*{Acknowledgments}
We thank P. Debenedetti and P. Poole for a critical reading of the manuscript.  L.R. acknowledges support from the Austrian Research Fund (FWF) through the Lise-Meitner Fellowship No. M 1650-N27 and from the European Commission through the Marie Sk\l{}odowska-Curie Fellowship No. 702298-
DELTAS V.B. acknowledges support from the Austrian Science Fund (FWF) project P 26253-N27. 
F.S. acknowledges support from ETN-COLLDENSE (H2020-MCSA-ITN-2014, Grant No. 642774). J.M.T. acknowledges financial support from the Portuguese Foundation for Science and Tecnhology under contracts EXCL/FIS-NAN/0083/2012 and UID/FIS/00618/2013.

\bibliography{./library}

\begin{thebibliography}{60}%
\makeatletter
\providecommand \@ifxundefined [1]{%
 \@ifx{#1\undefined}
}%
\providecommand \@ifnum [1]{%
 \ifnum #1\expandafter \@firstoftwo
 \else \expandafter \@secondoftwo
 \fi
}%
\providecommand \@ifx [1]{%
 \ifx #1\expandafter \@firstoftwo
 \else \expandafter \@secondoftwo
 \fi
}%
\providecommand \natexlab [1]{#1}%
\providecommand \enquote  [1]{``#1''}%
\providecommand \bibnamefont  [1]{#1}%
\providecommand \bibfnamefont [1]{#1}%
\providecommand \citenamefont [1]{#1}%
\providecommand \href@noop [0]{\@secondoftwo}%
\providecommand \href [0]{\begingroup \@sanitize@url \@href}%
\providecommand \@href[1]{\@@startlink{#1}\@@href}%
\providecommand \@@href[1]{\endgroup#1\@@endlink}%
\providecommand \@sanitize@url [0]{\catcode `\\12\catcode `\$12\catcode
  `\&12\catcode `\#12\catcode `\^12\catcode `\_12\catcode `\%12\relax}%
\providecommand \@@startlink[1]{}%
\providecommand \@@endlink[0]{}%
\providecommand \url  [0]{\begingroup\@sanitize@url \@url }%
\providecommand \@url [1]{\endgroup\@href {#1}{\urlprefix }}%
\providecommand \urlprefix  [0]{URL }%
\providecommand \Eprint [0]{\href }%
\providecommand \doibase [0]{http://dx.doi.org/}%
\providecommand \selectlanguage [0]{\@gobble}%
\providecommand \bibinfo  [0]{\@secondoftwo}%
\providecommand \bibfield  [0]{\@secondoftwo}%
\providecommand \translation [1]{[#1]}%
\providecommand \BibitemOpen [0]{}%
\providecommand \bibitemStop [0]{}%
\providecommand \bibitemNoStop [0]{.\EOS\space}%
\providecommand \EOS [0]{\spacefactor3000\relax}%
\providecommand \BibitemShut  [1]{\csname bibitem#1\endcsname}%
\let\auto@bib@innerbib\@empty
%</preamble>
\bibitem [{\citenamefont {Kanno}\ and\ \citenamefont
  {Angell}(1979)}]{kanno:4008}%
  \BibitemOpen
  \bibfield  {author} {\bibinfo {author} {\bibfnamefont {H.}~\bibnamefont
  {Kanno}}\ and\ \bibinfo {author} {\bibfnamefont {C.~A.}\ \bibnamefont
  {Angell}},\ }\href {\doibase 10.1063/1.438021} {\bibfield  {journal}
  {\bibinfo  {journal} {J. Chem. Phys.}\ }\textbf {\bibinfo {volume} {70}},\
  \bibinfo {pages} {4008} (\bibinfo {year} {1979})}\BibitemShut {NoStop}%
\bibitem [{\citenamefont {Angell}\ \emph {et~al.}(1982)\citenamefont {Angell},
  \citenamefont {Sichina},\ and\ \citenamefont
  {Oguni}}]{AngellC.A._j100395a032}%
  \BibitemOpen
  \bibfield  {author} {\bibinfo {author} {\bibfnamefont {C.~A.}\ \bibnamefont
  {Angell}}, \bibinfo {author} {\bibfnamefont {W.~J.}\ \bibnamefont {Sichina}},
  \ and\ \bibinfo {author} {\bibfnamefont {M.}~\bibnamefont {Oguni}},\ }\href
  {http://pubs3.acs.org/acs/journals/doilookup?in{\_}doi=10.1021/j100395a032}
  {\bibfield  {journal} {\bibinfo  {journal} {J. Phys. Chem.}\ }\textbf
  {\bibinfo {volume} {86}},\ \bibinfo {pages} {998} (\bibinfo {year}
  {1982})}\BibitemShut {NoStop}%
\bibitem [{\citenamefont {Debenedetti}(1996)}]{debenedetti-book}%
  \BibitemOpen
  \bibfield  {author} {\bibinfo {author} {\bibfnamefont {P.~G.}\ \bibnamefont
  {Debenedetti}},\ }\href@noop {} {\emph {\bibinfo {title} {{Metastable
  Liquids. Concepts and Principles}}}}\ (\bibinfo  {publisher} {Princeton
  University Press},\ \bibinfo {address} {Princeton, NJ},\ \bibinfo {year}
  {1996})\BibitemShut {NoStop}%
\bibitem [{\citenamefont {Poole}\ \emph {et~al.}(1994)\citenamefont {Poole},
  \citenamefont {Sciortino}, \citenamefont {Grande}, \citenamefont {Stanley},\
  and\ \citenamefont {Angell}}]{poolePRL1994}%
  \BibitemOpen
  \bibfield  {author} {\bibinfo {author} {\bibfnamefont {P.~H.}\ \bibnamefont
  {Poole}}, \bibinfo {author} {\bibfnamefont {F.}~\bibnamefont {Sciortino}},
  \bibinfo {author} {\bibfnamefont {T.}~\bibnamefont {Grande}}, \bibinfo
  {author} {\bibfnamefont {H.~E.}\ \bibnamefont {Stanley}}, \ and\ \bibinfo
  {author} {\bibfnamefont {C.~A.}\ \bibnamefont {Angell}},\ }\href {\doibase
  10.1103/PhysRevLett.73.1632} {\bibfield  {journal} {\bibinfo  {journal}
  {Phys. Rev. Lett.}\ }\textbf {\bibinfo {volume} {73}},\ \bibinfo {pages}
  {1632} (\bibinfo {year} {1994})}\BibitemShut {NoStop}%
\bibitem [{\citenamefont {Tanaka}(1996)}]{Tanaka96}%
  \BibitemOpen
  \bibfield  {author} {\bibinfo {author} {\bibfnamefont {H.}~\bibnamefont
  {Tanaka}},\ }\href@noop {} {\bibfield  {journal} {\bibinfo  {journal}
  {Nature}\ }\textbf {\bibinfo {volume} {380}},\ \bibinfo {pages} {328}
  (\bibinfo {year} {1996})}\BibitemShut {NoStop}%
\bibitem [{\citenamefont {Mishima}\ and\ \citenamefont
  {Stanley}(1998)}]{Mishima1998}%
  \BibitemOpen
  \bibfield  {author} {\bibinfo {author} {\bibfnamefont {O.}~\bibnamefont
  {Mishima}}\ and\ \bibinfo {author} {\bibfnamefont {H.~E.}\ \bibnamefont
  {Stanley}},\ }\href {\doibase 10.1038/32386} {\bibfield  {journal} {\bibinfo
  {journal} {Nature}\ }\textbf {\bibinfo {volume} {392}},\ \bibinfo {pages}
  {164} (\bibinfo {year} {1998})}\BibitemShut {NoStop}%
\bibitem [{\citenamefont {Soper}\ and\ \citenamefont
  {Ricci}(2000)}]{SoperPRL2000}%
  \BibitemOpen
  \bibfield  {author} {\bibinfo {author} {\bibfnamefont {A.~K.}\ \bibnamefont
  {Soper}}\ and\ \bibinfo {author} {\bibfnamefont {M.~A.}\ \bibnamefont
  {Ricci}},\ }\href@noop {} {\bibfield  {journal} {\bibinfo  {journal} {Phys.
  Rev. Lett.}\ }\textbf {\bibinfo {volume} {84}},\ \bibinfo {pages} {2881}
  (\bibinfo {year} {2000})}\BibitemShut {NoStop}%
\bibitem [{\citenamefont {Xu}\ \emph {et~al.}(2005)\citenamefont {Xu},
  \citenamefont {Kumar}, \citenamefont {Buldyrev}, \citenamefont {Chen},
  \citenamefont {Poole}, \citenamefont {Sciortino},\ and\ \citenamefont
  {Stanley}}]{Xu2005}%
  \BibitemOpen
  \bibfield  {author} {\bibinfo {author} {\bibfnamefont {L.}~\bibnamefont
  {Xu}}, \bibinfo {author} {\bibfnamefont {P.}~\bibnamefont {Kumar}}, \bibinfo
  {author} {\bibfnamefont {S.~V.}\ \bibnamefont {Buldyrev}}, \bibinfo {author}
  {\bibfnamefont {S.-H.}\ \bibnamefont {Chen}}, \bibinfo {author}
  {\bibfnamefont {P.~H.}\ \bibnamefont {Poole}}, \bibinfo {author}
  {\bibfnamefont {F.}~\bibnamefont {Sciortino}}, \ and\ \bibinfo {author}
  {\bibfnamefont {H.~E.}\ \bibnamefont {Stanley}},\ }\href {\doibase
  10.1073/pnas.0507870102} {\bibfield  {journal} {\bibinfo  {journal} {Proc.
  Natl. Acad. Sci. U.S.A.}\ }\textbf {\bibinfo {volume} {102}},\ \bibinfo
  {pages} {16558} (\bibinfo {year} {2005})}\BibitemShut {NoStop}%
\bibitem [{\citenamefont {Liu}\ \emph {et~al.}(2007)\citenamefont {Liu},
  \citenamefont {Zhang}, \citenamefont {Chen}, \citenamefont {Mou},
  \citenamefont {Poole},\ and\ \citenamefont {Chen}}]{LiuPNAS2007}%
  \BibitemOpen
  \bibfield  {author} {\bibinfo {author} {\bibfnamefont {D.}~\bibnamefont
  {Liu}}, \bibinfo {author} {\bibfnamefont {Y.}~\bibnamefont {Zhang}}, \bibinfo
  {author} {\bibfnamefont {C.-C.}\ \bibnamefont {Chen}}, \bibinfo {author}
  {\bibfnamefont {C.-Y.}\ \bibnamefont {Mou}}, \bibinfo {author} {\bibfnamefont
  {P.~H.}\ \bibnamefont {Poole}}, \ and\ \bibinfo {author} {\bibfnamefont
  {S.-H.}\ \bibnamefont {Chen}},\ }\href {\doibase 10.1073/pnas.0701352104}
  {\bibfield  {journal} {\bibinfo  {journal} {Proc. Natl. Acad. Sci. U.S.A.}\
  }\textbf {\bibinfo {volume} {104}},\ \bibinfo {pages} {9570} (\bibinfo {year}
  {2007})}\BibitemShut {NoStop}%
\bibitem [{\citenamefont {Stokely}\ \emph {et~al.}(2010)\citenamefont
  {Stokely}, \citenamefont {Mazza}, \citenamefont {Stanley},\ and\
  \citenamefont {Franzese}}]{StokelyPNAS2010}%
  \BibitemOpen
  \bibfield  {author} {\bibinfo {author} {\bibfnamefont {K.}~\bibnamefont
  {Stokely}}, \bibinfo {author} {\bibfnamefont {M.~G.}\ \bibnamefont {Mazza}},
  \bibinfo {author} {\bibfnamefont {H.~E.}\ \bibnamefont {Stanley}}, \ and\
  \bibinfo {author} {\bibfnamefont {G.}~\bibnamefont {Franzese}},\ }\href@noop
  {} {\bibfield  {journal} {\bibinfo  {journal} {Proc. Natl. Acad. Sci.
  U.S.A.}\ }\textbf {\bibinfo {volume} {107}},\ \bibinfo {pages} {1301}
  (\bibinfo {year} {2010})}\BibitemShut {NoStop}%
\bibitem [{\citenamefont {Abascal}\ and\ \citenamefont
  {Vega}(2010)}]{abascalJCP2010}%
  \BibitemOpen
  \bibfield  {author} {\bibinfo {author} {\bibfnamefont {J.~L.~F.}\
  \bibnamefont {Abascal}}\ and\ \bibinfo {author} {\bibfnamefont
  {C.}~\bibnamefont {Vega}},\ }\href {\doibase 10.1063/1.3506860} {\bibfield
  {journal} {\bibinfo  {journal} {J. Chem. Phys.}\ }\textbf {\bibinfo {volume}
  {133}},\ \bibinfo {pages} {234502} (\bibinfo {year} {2010})}\BibitemShut
  {NoStop}%
\bibitem [{\citenamefont {Bertrand}\ and\ \citenamefont
  {Anisimov}(2011)}]{bertrand2011}%
  \BibitemOpen
  \bibfield  {author} {\bibinfo {author} {\bibfnamefont {C.~E.}\ \bibnamefont
  {Bertrand}}\ and\ \bibinfo {author} {\bibfnamefont {M.~A.}\ \bibnamefont
  {Anisimov}},\ }\href {\doibase 10.1021/jp204011z} {\bibfield  {journal}
  {\bibinfo  {journal} {J. Phys. Chem. B}\ }\textbf {\bibinfo {volume} {115}},\
  \bibinfo {pages} {14099} (\bibinfo {year} {2011})}\BibitemShut {NoStop}%
\bibitem [{\citenamefont {Holten}\ \emph {et~al.}(2013)\citenamefont {Holten},
  \citenamefont {Limmer}, \citenamefont {Molinero},\ and\ \citenamefont
  {Anisimov}}]{holtenJCP2013}%
  \BibitemOpen
  \bibfield  {author} {\bibinfo {author} {\bibfnamefont {V.}~\bibnamefont
  {Holten}}, \bibinfo {author} {\bibfnamefont {D.~T.}\ \bibnamefont {Limmer}},
  \bibinfo {author} {\bibfnamefont {V.}~\bibnamefont {Molinero}}, \ and\
  \bibinfo {author} {\bibfnamefont {M.~A.}\ \bibnamefont {Anisimov}},\ }\href
  {\doibase 10.1063/1.4802992} {\bibfield  {journal} {\bibinfo  {journal} {J.
  Chem. Phys.}\ }\textbf {\bibinfo {volume} {138}},\ \bibinfo {pages} {174501}
  (\bibinfo {year} {2013})}\BibitemShut {NoStop}%
\bibitem [{\citenamefont {Azouzi}\ \emph {et~al.}(2013)\citenamefont {Azouzi},
  \citenamefont {Ramboz}, \citenamefont {Lenain},\ and\ \citenamefont
  {Caupin}}]{azouzi2013coherent}%
  \BibitemOpen
  \bibfield  {author} {\bibinfo {author} {\bibfnamefont {M.~E.~M.}\
  \bibnamefont {Azouzi}}, \bibinfo {author} {\bibfnamefont {C.}~\bibnamefont
  {Ramboz}}, \bibinfo {author} {\bibfnamefont {J.-F.}\ \bibnamefont {Lenain}},
  \ and\ \bibinfo {author} {\bibfnamefont {F.}~\bibnamefont {Caupin}},\
  }\href@noop {} {\bibfield  {journal} {\bibinfo  {journal} {Nature Physics}\
  }\textbf {\bibinfo {volume} {9}},\ \bibinfo {pages} {38} (\bibinfo {year}
  {2013})}\BibitemShut {NoStop}%
\bibitem [{\citenamefont {Pallares}\ \emph {et~al.}(2014)\citenamefont
  {Pallares}, \citenamefont {Azouzi}, \citenamefont {Gonz{\'a}lez},
  \citenamefont {Aragones}, \citenamefont {Abascal}, \citenamefont
  {Valeriani},\ and\ \citenamefont {Caupin}}]{pallares2014anomalies}%
  \BibitemOpen
  \bibfield  {author} {\bibinfo {author} {\bibfnamefont {G.}~\bibnamefont
  {Pallares}}, \bibinfo {author} {\bibfnamefont {M.~E.~M.}\ \bibnamefont
  {Azouzi}}, \bibinfo {author} {\bibfnamefont {M.~A.}\ \bibnamefont
  {Gonz{\'a}lez}}, \bibinfo {author} {\bibfnamefont {J.~L.}\ \bibnamefont
  {Aragones}}, \bibinfo {author} {\bibfnamefont {J.~L.}\ \bibnamefont
  {Abascal}}, \bibinfo {author} {\bibfnamefont {C.}~\bibnamefont {Valeriani}},
  \ and\ \bibinfo {author} {\bibfnamefont {F.}~\bibnamefont {Caupin}},\
  }\href@noop {} {\bibfield  {journal} {\bibinfo  {journal} {Proceedings of the
  National Academy of Sciences}\ }\textbf {\bibinfo {volume} {111}},\ \bibinfo
  {pages} {7936} (\bibinfo {year} {2014})}\BibitemShut {NoStop}%
\bibitem [{\citenamefont {Bianco}\ and\ \citenamefont
  {Franzese}(2014)}]{BiancoSR2014}%
  \BibitemOpen
  \bibfield  {author} {\bibinfo {author} {\bibfnamefont {V.}~\bibnamefont
  {Bianco}}\ and\ \bibinfo {author} {\bibfnamefont {G.}~\bibnamefont
  {Franzese}},\ }\href
  {http://www.pubmedcentral.nih.gov/articlerender.fcgi?artid=3975237{\&}tool=pmcentrez{\&}rendertype=abstract}
  {\bibfield  {journal} {\bibinfo  {journal} {Scientific Reports}\ }\textbf
  {\bibinfo {volume} {4}},\ \bibinfo {pages} {4440} (\bibinfo {year}
  {2014})}\BibitemShut {NoStop}%
\bibitem [{\citenamefont {Soper}(2014)}]{Soper2014}%
  \BibitemOpen
  \bibfield  {author} {\bibinfo {author} {\bibfnamefont {A.~K.}\ \bibnamefont
  {Soper}},\ }\href {\doibase 10.1038/nmat4019} {\bibfield  {journal} {\bibinfo
   {journal} {Nature materials}\ }\textbf {\bibinfo {volume} {13}},\ \bibinfo
  {pages} {671} (\bibinfo {year} {2014})}\BibitemShut {NoStop}%
\bibitem [{\citenamefont {Nilsson}\ and\ \citenamefont
  {Pettersson}(2015)}]{Nilsson2015}%
  \BibitemOpen
  \bibfield  {author} {\bibinfo {author} {\bibfnamefont {A.}~\bibnamefont
  {Nilsson}}\ and\ \bibinfo {author} {\bibfnamefont {L.~G.~M.}\ \bibnamefont
  {Pettersson}},\ }\href {\doibase 10.1038/ncomms9998} {\bibfield  {journal}
  {\bibinfo  {journal} {Nature communications}\ }\textbf {\bibinfo {volume}
  {6}},\ \bibinfo {pages} {8998} (\bibinfo {year} {2015})}\BibitemShut
  {NoStop}%
\bibitem [{\citenamefont {Speedy}(1982{\natexlab{a}})}]{Speedy1982}%
  \BibitemOpen
  \bibfield  {author} {\bibinfo {author} {\bibfnamefont {R.~J.}\ \bibnamefont
  {Speedy}},\ }\href {\doibase 10.1021/j100395a030} {\bibfield  {journal}
  {\bibinfo  {journal} {J. Phys. Chem.}\ }\textbf {\bibinfo {volume} {86}},\
  \bibinfo {pages} {982} (\bibinfo {year} {1982}{\natexlab{a}})}\BibitemShut
  {NoStop}%
\bibitem [{\citenamefont {Sastry}\ \emph {et~al.}(1996)\citenamefont {Sastry},
  \citenamefont {Debenedetti}, \citenamefont {Sciortino},\ and\ \citenamefont
  {Stanley}}]{Sastry1996}%
  \BibitemOpen
  \bibfield  {author} {\bibinfo {author} {\bibfnamefont {S.}~\bibnamefont
  {Sastry}}, \bibinfo {author} {\bibfnamefont {P.~G.}\ \bibnamefont
  {Debenedetti}}, \bibinfo {author} {\bibfnamefont {F.}~\bibnamefont
  {Sciortino}}, \ and\ \bibinfo {author} {\bibfnamefont {H.~E.}\ \bibnamefont
  {Stanley}},\ }\href {\doibase 10.1103/PhysRevE.53.6144} {\bibfield  {journal}
  {\bibinfo  {journal} {Phys. Rev. E}\ }\textbf {\bibinfo {volume} {53}},\
  \bibinfo {pages} {6144} (\bibinfo {year} {1996})}\BibitemShut {NoStop}%
\bibitem [{\citenamefont {Poole}\ \emph {et~al.}(1992)\citenamefont {Poole},
  \citenamefont {Sciortino}, \citenamefont {Essmann},\ and\ \citenamefont
  {Stanley}}]{llcp}%
  \BibitemOpen
  \bibfield  {author} {\bibinfo {author} {\bibfnamefont {P.~H.}\ \bibnamefont
  {Poole}}, \bibinfo {author} {\bibfnamefont {F.}~\bibnamefont {Sciortino}},
  \bibinfo {author} {\bibfnamefont {U.}~\bibnamefont {Essmann}}, \ and\
  \bibinfo {author} {\bibfnamefont {H.~E.}\ \bibnamefont {Stanley}},\
  }\href@noop {} {\bibfield  {journal} {\bibinfo  {journal} {Nature}\ }\textbf
  {\bibinfo {volume} {360}},\ \bibinfo {pages} {324} (\bibinfo {year}
  {1992})}\BibitemShut {NoStop}%
\bibitem [{\citenamefont {Angell}(2008)}]{AngellScience2008}%
  \BibitemOpen
  \bibfield  {author} {\bibinfo {author} {\bibfnamefont {C.~A.}\ \bibnamefont
  {Angell}},\ }\href {\doibase 10.1126/science.1131939} {\bibfield  {journal}
  {\bibinfo  {journal} {Science}\ }\textbf {\bibinfo {volume} {319}},\ \bibinfo
  {pages} {582} (\bibinfo {year} {2008})}\BibitemShut {NoStop}%
\bibitem [{\citenamefont {Sciortino}\ \emph {et~al.}(2011)\citenamefont
  {Sciortino}, \citenamefont {Saika-Voivod},\ and\ \citenamefont
  {Poole}}]{sciortinoPCCP2011}%
  \BibitemOpen
  \bibfield  {author} {\bibinfo {author} {\bibfnamefont {F.}~\bibnamefont
  {Sciortino}}, \bibinfo {author} {\bibfnamefont {I.}~\bibnamefont
  {Saika-Voivod}}, \ and\ \bibinfo {author} {\bibfnamefont {P.~H.}\
  \bibnamefont {Poole}},\ }\href {\doibase 10.1039/C1CP22316J} {\bibfield
  {journal} {\bibinfo  {journal} {Phys. Chem. Chem. Phys.}\ }\textbf {\bibinfo
  {volume} {13}},\ \bibinfo {pages} {19759} (\bibinfo {year}
  {2011})}\BibitemShut {NoStop}%
\bibitem [{\citenamefont {Liu}\ \emph {et~al.}(2012)\citenamefont {Liu},
  \citenamefont {Palmer}, \citenamefont {Panagiotopoulos},\ and\ \citenamefont
  {Debenedetti}}]{liuJCP2012}%
  \BibitemOpen
  \bibfield  {author} {\bibinfo {author} {\bibfnamefont {Y.}~\bibnamefont
  {Liu}}, \bibinfo {author} {\bibfnamefont {J.~C.}\ \bibnamefont {Palmer}},
  \bibinfo {author} {\bibfnamefont {A.~Z.}\ \bibnamefont {Panagiotopoulos}}, \
  and\ \bibinfo {author} {\bibfnamefont {P.~G.}\ \bibnamefont {Debenedetti}},\
  }\href {\doibase 10.1063/1.4769126} {\bibfield  {journal} {\bibinfo
  {journal} {J. Chem. Phys.}\ }\textbf {\bibinfo {volume} {137}},\ \bibinfo
  {pages} {214505} (\bibinfo {year} {2012})}\BibitemShut {NoStop}%
\bibitem [{\citenamefont {Palmer}\ \emph {et~al.}(2013)\citenamefont {Palmer},
  \citenamefont {Car},\ and\ \citenamefont
  {Debenedetti}}]{palmer_car_debenedetti}%
  \BibitemOpen
  \bibfield  {author} {\bibinfo {author} {\bibfnamefont {J.~C.}\ \bibnamefont
  {Palmer}}, \bibinfo {author} {\bibfnamefont {R.}~\bibnamefont {Car}}, \ and\
  \bibinfo {author} {\bibfnamefont {P.~G.}\ \bibnamefont {Debenedetti}},\
  }\href {\doibase 10.1039/C3FD00074E} {\bibfield  {journal} {\bibinfo
  {journal} {Faraday Discuss.}\ }\textbf {\bibinfo {volume} {167}},\ \bibinfo
  {pages} {77} (\bibinfo {year} {2013})}\BibitemShut {NoStop}%
\bibitem [{\citenamefont {Poole}\ \emph {et~al.}(2013)\citenamefont {Poole},
  \citenamefont {Bowles}, \citenamefont {Saika-Voivod},\ and\ \citenamefont
  {Sciortino}}]{pooleJCP2013}%
  \BibitemOpen
  \bibfield  {author} {\bibinfo {author} {\bibfnamefont {P.~H.}\ \bibnamefont
  {Poole}}, \bibinfo {author} {\bibfnamefont {R.~K.}\ \bibnamefont {Bowles}},
  \bibinfo {author} {\bibfnamefont {I.}~\bibnamefont {Saika-Voivod}}, \ and\
  \bibinfo {author} {\bibfnamefont {F.}~\bibnamefont {Sciortino}},\ }\href
  {\doibase 10.1063/1.4775738} {\bibfield  {journal} {\bibinfo  {journal} {J.
  Chem. Phys.}\ }\textbf {\bibinfo {volume} {138}},\ \bibinfo {pages} {34505}
  (\bibinfo {year} {2013})}\BibitemShut {NoStop}%
\bibitem [{\citenamefont {Palmer}\ \emph {et~al.}(2014)\citenamefont {Palmer},
  \citenamefont {Martelli}, \citenamefont {Liu}, \citenamefont {Car},
  \citenamefont {Panagiotopoulos},\ and\ \citenamefont
  {Debenedetti}}]{Palmer2014}%
  \BibitemOpen
  \bibfield  {author} {\bibinfo {author} {\bibfnamefont {J.~C.}\ \bibnamefont
  {Palmer}}, \bibinfo {author} {\bibfnamefont {F.}~\bibnamefont {Martelli}},
  \bibinfo {author} {\bibfnamefont {Y.}~\bibnamefont {Liu}}, \bibinfo {author}
  {\bibfnamefont {R.}~\bibnamefont {Car}}, \bibinfo {author} {\bibfnamefont
  {A.~Z.}\ \bibnamefont {Panagiotopoulos}}, \ and\ \bibinfo {author}
  {\bibfnamefont {P.~G.}\ \bibnamefont {Debenedetti}},\ }\href {\doibase
  10.1038/nature13405} {\bibfield  {journal} {\bibinfo  {journal} {Nature}\
  }\textbf {\bibinfo {volume} {510}},\ \bibinfo {pages} {385} (\bibinfo {year}
  {2014})}\BibitemShut {NoStop}%
\bibitem [{\citenamefont {Smallenburg}\ and\ \citenamefont
  {Sciortino}(2015)}]{SmallenburgPRL2015}%
  \BibitemOpen
  \bibfield  {author} {\bibinfo {author} {\bibfnamefont {F.}~\bibnamefont
  {Smallenburg}}\ and\ \bibinfo {author} {\bibfnamefont {F.}~\bibnamefont
  {Sciortino}},\ }\href {\doibase 10.1103/PhysRevLett.115.015701} {\bibfield
  {journal} {\bibinfo  {journal} {Phys. Rev. Lett.}\ }\textbf {\bibinfo
  {volume} {115}},\ \bibinfo {pages} {015701} (\bibinfo {year}
  {2015})}\BibitemShut {NoStop}%
\bibitem [{\citenamefont {Speedy}(1982{\natexlab{b}})}]{Speedy82}%
  \BibitemOpen
  \bibfield  {author} {\bibinfo {author} {\bibfnamefont {R.~J.}\ \bibnamefont
  {Speedy}},\ }\href {http://dx.doi.org/10.1021/j100212a038} {\bibfield
  {journal} {\bibinfo  {journal} {J. Phys. Chem.}\ }\textbf {\bibinfo {volume}
  {86}},\ \bibinfo {pages} {3002} (\bibinfo {year}
  {1982}{\natexlab{b}})}\BibitemShut {NoStop}%
\bibitem [{\citenamefont {Poole}\ \emph {et~al.}(1993)\citenamefont {Poole},
  \citenamefont {Sciortino}, \citenamefont {Essmann},\ and\ \citenamefont
  {Stanley}}]{poolePRE1993}%
  \BibitemOpen
  \bibfield  {author} {\bibinfo {author} {\bibfnamefont {P.~H.}\ \bibnamefont
  {Poole}}, \bibinfo {author} {\bibfnamefont {F.}~\bibnamefont {Sciortino}},
  \bibinfo {author} {\bibfnamefont {U.}~\bibnamefont {Essmann}}, \ and\
  \bibinfo {author} {\bibfnamefont {H.~E.}\ \bibnamefont {Stanley}},\ }\href
  {\doibase 10.1103/PhysRevE.48.3799} {\bibfield  {journal} {\bibinfo
  {journal} {Phys. Rev. E}\ }\textbf {\bibinfo {volume} {48}},\ \bibinfo
  {pages} {3799} (\bibinfo {year} {1993})}\BibitemShut {NoStop}%
\bibitem [{\citenamefont {Debenedetti}(2003)}]{De03}%
  \BibitemOpen
  \bibfield  {author} {\bibinfo {author} {\bibfnamefont {P.~G.}\ \bibnamefont
  {Debenedetti}},\ }\href {http://stacks.iop.org/0953-8984/15/R1669} {\bibfield
   {journal} {\bibinfo  {journal} {Journal of Physics: Condensed Matter}\
  }\textbf {\bibinfo {volume} {15}},\ \bibinfo {pages} {R1669} (\bibinfo {year}
  {2003})}\BibitemShut {NoStop}%
\bibitem [{\citenamefont {Sastry}\ \emph {et~al.}(1993)\citenamefont {Sastry},
  \citenamefont {Sciortino},\ and\ \citenamefont {Stanley}}]{Sastry1993}%
  \BibitemOpen
  \bibfield  {author} {\bibinfo {author} {\bibfnamefont {S.}~\bibnamefont
  {Sastry}}, \bibinfo {author} {\bibfnamefont {F.}~\bibnamefont {Sciortino}}, \
  and\ \bibinfo {author} {\bibfnamefont {H.}~\bibnamefont {Stanley}},\ }\href
  {\doibase 10.1016/0009-2614(93)87026-Y} {\bibfield  {journal} {\bibinfo
  {journal} {Chemical Physics Letters}\ }\textbf {\bibinfo {volume} {207}},\
  \bibinfo {pages} {275} (\bibinfo {year} {1993})}\BibitemShut {NoStop}%
\bibitem [{\citenamefont {Sasai}(1993)}]{Sasai1993}%
  \BibitemOpen
  \bibfield  {author} {\bibinfo {author} {\bibfnamefont {M.}~\bibnamefont
  {Sasai}},\ }\href {\doibase 10.1246/bcsj.66.3362} {\bibfield  {journal}
  {\bibinfo  {journal} {Bulletin of the Chemical Society of Japan}\ }\textbf
  {\bibinfo {volume} {66}},\ \bibinfo {pages} {3362} (\bibinfo {year}
  {1993})}\BibitemShut {NoStop}%
\bibitem [{\citenamefont {Borick}\ \emph {et~al.}(1995)\citenamefont {Borick},
  \citenamefont {Debenedetti},\ and\ \citenamefont {Sastry}}]{Borick1995}%
  \BibitemOpen
  \bibfield  {author} {\bibinfo {author} {\bibfnamefont {S.~S.}\ \bibnamefont
  {Borick}}, \bibinfo {author} {\bibfnamefont {P.~G.}\ \bibnamefont
  {Debenedetti}}, \ and\ \bibinfo {author} {\bibfnamefont {S.}~\bibnamefont
  {Sastry}},\ }\href {\doibase 10.1021/j100011a054} {\bibfield  {journal}
  {\bibinfo  {journal} {J. Phys. Chem.}\ }\textbf {\bibinfo {volume} {99}},\
  \bibinfo {pages} {3781} (\bibinfo {year} {1995})}\BibitemShut {NoStop}%
\bibitem [{\citenamefont {Kern}\ and\ \citenamefont
  {Frenkel}(2003)}]{Kern2003}%
  \BibitemOpen
  \bibfield  {author} {\bibinfo {author} {\bibfnamefont {N.}~\bibnamefont
  {Kern}}\ and\ \bibinfo {author} {\bibfnamefont {D.}~\bibnamefont {Frenkel}},\
  }\href {\doibase 10.1063/1.1569473} {\bibfield  {journal} {\bibinfo
  {journal} {J. Chem. Phys.}\ }\textbf {\bibinfo {volume} {118}},\ \bibinfo
  {pages} {9882} (\bibinfo {year} {2003})}\BibitemShut {NoStop}%
\bibitem [{\citenamefont {Bianchi}\ \emph {et~al.}(2011)\citenamefont
  {Bianchi}, \citenamefont {Blaak},\ and\ \citenamefont {Likos}}]{Bianchi2011}%
  \BibitemOpen
  \bibfield  {author} {\bibinfo {author} {\bibfnamefont {E.}~\bibnamefont
  {Bianchi}}, \bibinfo {author} {\bibfnamefont {R.}~\bibnamefont {Blaak}}, \
  and\ \bibinfo {author} {\bibfnamefont {C.~N.}\ \bibnamefont {Likos}},\ }\href
  {\doibase 10.1039/c0cp02296a} {\bibfield  {journal} {\bibinfo  {journal}
  {Physical Chemistry Chemical Physics : PCCP}\ }\textbf {\bibinfo {volume}
  {13}},\ \bibinfo {pages} {6397} (\bibinfo {year} {2011})}\BibitemShut
  {NoStop}%
\bibitem [{\citenamefont {Bianchi}\ \emph {et~al.}(2006)\citenamefont
  {Bianchi}, \citenamefont {Largo}, \citenamefont {Tartaglia}, \citenamefont
  {Zaccarelli},\ and\ \citenamefont {Sciortino}}]{Bianchi2006}%
  \BibitemOpen
  \bibfield  {author} {\bibinfo {author} {\bibfnamefont {E.}~\bibnamefont
  {Bianchi}}, \bibinfo {author} {\bibfnamefont {J.}~\bibnamefont {Largo}},
  \bibinfo {author} {\bibfnamefont {P.}~\bibnamefont {Tartaglia}}, \bibinfo
  {author} {\bibfnamefont {E.}~\bibnamefont {Zaccarelli}}, \ and\ \bibinfo
  {author} {\bibfnamefont {F.}~\bibnamefont {Sciortino}},\ }\href {\doibase
  10.1103/PhysRevLett.97.168301} {\bibfield  {journal} {\bibinfo  {journal}
  {Phys. Rev. Lett.}\ }\textbf {\bibinfo {volume} {97}},\ \bibinfo {pages} {4}
  (\bibinfo {year} {2006})}\BibitemShut {NoStop}%
\bibitem [{\citenamefont {Romano}\ \emph {et~al.}(2011)\citenamefont {Romano},
  \citenamefont {Sanz},\ and\ \citenamefont {Sciortino}}]{flavio_tetra}%
  \BibitemOpen
  \bibfield  {author} {\bibinfo {author} {\bibfnamefont {F.}~\bibnamefont
  {Romano}}, \bibinfo {author} {\bibfnamefont {E.}~\bibnamefont {Sanz}}, \ and\
  \bibinfo {author} {\bibfnamefont {F.}~\bibnamefont {Sciortino}},\ }\href
  {\doibase 10.1063/1.3578182} {\bibfield  {journal} {\bibinfo  {journal} {J.
  Chem. Phys.}\ }\textbf {\bibinfo {volume} {134}},\ \bibinfo {pages} {174502}
  (\bibinfo {year} {2011})}\BibitemShut {NoStop}%
\bibitem [{\citenamefont {Doppelbauer}\ \emph {et~al.}(2012)\citenamefont
  {Doppelbauer}, \citenamefont {Noya}, \citenamefont {Bianchi},\ and\
  \citenamefont {Kahl}}]{doppelbauer_patchy_genetic}%
  \BibitemOpen
  \bibfield  {author} {\bibinfo {author} {\bibfnamefont {G.}~\bibnamefont
  {Doppelbauer}}, \bibinfo {author} {\bibfnamefont {E.~G.}\ \bibnamefont
  {Noya}}, \bibinfo {author} {\bibfnamefont {E.}~\bibnamefont {Bianchi}}, \
  and\ \bibinfo {author} {\bibfnamefont {G.}~\bibnamefont {Kahl}},\ }\href
  {\doibase 10.1039/C2SM26043C} {\bibfield  {journal} {\bibinfo  {journal}
  {Soft Matter}\ }\textbf {\bibinfo {volume} {8}},\ \bibinfo {pages} {7768}
  (\bibinfo {year} {2012})}\BibitemShut {NoStop}%
\bibitem [{\citenamefont {Smallenburg}\ \emph {et~al.}(2014)\citenamefont
  {Smallenburg}, \citenamefont {Filion},\ and\ \citenamefont
  {Sciortino}}]{smallenburg2014erasing}%
  \BibitemOpen
  \bibfield  {author} {\bibinfo {author} {\bibfnamefont {F.}~\bibnamefont
  {Smallenburg}}, \bibinfo {author} {\bibfnamefont {L.}~\bibnamefont {Filion}},
  \ and\ \bibinfo {author} {\bibfnamefont {F.}~\bibnamefont {Sciortino}},\
  }\href@noop {} {\bibfield  {journal} {\bibinfo  {journal} {Nature Physics}\
  }\textbf {\bibinfo {volume} {10}},\ \bibinfo {pages} {653} (\bibinfo {year}
  {2014})}\BibitemShut {NoStop}%
\bibitem [{\citenamefont {Wilber}\ \emph {et~al.}(2009)\citenamefont {Wilber},
  \citenamefont {Doye}, \citenamefont {Louis},\ and\ \citenamefont
  {Lewis}}]{Wilber2009}%
  \BibitemOpen
  \bibfield  {author} {\bibinfo {author} {\bibfnamefont {A.~W.}\ \bibnamefont
  {Wilber}}, \bibinfo {author} {\bibfnamefont {J.~P.~K.}\ \bibnamefont {Doye}},
  \bibinfo {author} {\bibfnamefont {A.~A.}\ \bibnamefont {Louis}}, \ and\
  \bibinfo {author} {\bibfnamefont {A.~C.~F.}\ \bibnamefont {Lewis}},\ }\href
  {\doibase 10.1063/1.3243581} {\bibfield  {journal} {\bibinfo  {journal} {J.
  Chem. Phys.}\ }\textbf {\bibinfo {volume} {131}},\ \bibinfo {pages} {175102}
  (\bibinfo {year} {2009})}\BibitemShut {NoStop}%
\bibitem [{\citenamefont {de~las Heras}\ \emph {et~al.}(2012)\citenamefont
  {de~las Heras}, \citenamefont {Tavares},\ and\ \citenamefont
  {da~Gama}}]{delasheras_bicontinuous_gels}%
  \BibitemOpen
  \bibfield  {author} {\bibinfo {author} {\bibfnamefont {D.}~\bibnamefont
  {de~las Heras}}, \bibinfo {author} {\bibfnamefont {J.~M.}\ \bibnamefont
  {Tavares}}, \ and\ \bibinfo {author} {\bibfnamefont {M.~M.}\ \bibnamefont
  {da~Gama}},\ }\href {\doibase 10.1039/C1SM06948A} {\bibfield  {journal}
  {\bibinfo  {journal} {Soft Matter}\ }\textbf {\bibinfo {volume} {8}},\
  \bibinfo {pages} {1785} (\bibinfo {year} {2012})}\BibitemShut {NoStop}%
\bibitem [{\citenamefont {Coluzza}\ \emph {et~al.}(2013)\citenamefont
  {Coluzza}, \citenamefont {van Oostrum}, \citenamefont {Capone}, \citenamefont
  {Reimhult},\ and\ \citenamefont {Dellago}}]{Coluzza2013}%
  \BibitemOpen
  \bibfield  {author} {\bibinfo {author} {\bibfnamefont {I.}~\bibnamefont
  {Coluzza}}, \bibinfo {author} {\bibfnamefont {P.}~\bibnamefont {van
  Oostrum}}, \bibinfo {author} {\bibfnamefont {B.}~\bibnamefont {Capone}},
  \bibinfo {author} {\bibfnamefont {E.}~\bibnamefont {Reimhult}}, \ and\
  \bibinfo {author} {\bibfnamefont {C.}~\bibnamefont {Dellago}},\ }\href
  {\doibase 10.1103/PhysRevLett.110.075501} {\bibfield  {journal} {\bibinfo
  {journal} {Phys. Rev. Lett.}\ }\textbf {\bibinfo {volume} {110}},\ \bibinfo
  {pages} {75501} (\bibinfo {year} {2013})}\BibitemShut {NoStop}%
\bibitem [{\citenamefont {Hong}\ \emph {et~al.}(2008)\citenamefont {Hong},
  \citenamefont {Cacciuto}, \citenamefont {Luijten},\ and\ \citenamefont
  {Granick}}]{Hong2008}%
  \BibitemOpen
  \bibfield  {author} {\bibinfo {author} {\bibfnamefont {L.}~\bibnamefont
  {Hong}}, \bibinfo {author} {\bibfnamefont {A.}~\bibnamefont {Cacciuto}},
  \bibinfo {author} {\bibfnamefont {E.}~\bibnamefont {Luijten}}, \ and\
  \bibinfo {author} {\bibfnamefont {S.}~\bibnamefont {Granick}},\ }\href
  {http://pubs.acs.org/cgi-bin/abstract.cgi/langd5/2008/24/i03/abs/la7030818.html
  papers2://publication/doi/10.1021/la7030818} {\bibfield  {journal} {\bibinfo
  {journal} {Langmuir}\ }\textbf {\bibinfo {volume} {24}},\ \bibinfo {pages}
  {621} (\bibinfo {year} {2008})}\BibitemShut {NoStop}%
\bibitem [{\citenamefont {Sciortino}\ \emph {et~al.}(2009)\citenamefont
  {Sciortino}, \citenamefont {Giacometti},\ and\ \citenamefont
  {Pastore}}]{janus_pd}%
  \BibitemOpen
  \bibfield  {author} {\bibinfo {author} {\bibfnamefont {F.}~\bibnamefont
  {Sciortino}}, \bibinfo {author} {\bibfnamefont {A.}~\bibnamefont
  {Giacometti}}, \ and\ \bibinfo {author} {\bibfnamefont {G.}~\bibnamefont
  {Pastore}},\ }\href {\doibase 10.1103/PhysRevLett.103.237801} {\bibfield
  {journal} {\bibinfo  {journal} {Phys. Rev. Lett.}\ }\textbf {\bibinfo
  {volume} {103}},\ \bibinfo {pages} {237801} (\bibinfo {year}
  {2009})}\BibitemShut {NoStop}%
\bibitem [{\citenamefont {Vissers}\ \emph {et~al.}(2013)\citenamefont
  {Vissers}, \citenamefont {Preisler}, \citenamefont {Smallenburg},
  \citenamefont {Dijkstra},\ and\ \citenamefont {Sciortino}}]{janus_crystals}%
  \BibitemOpen
  \bibfield  {author} {\bibinfo {author} {\bibfnamefont {T.}~\bibnamefont
  {Vissers}}, \bibinfo {author} {\bibfnamefont {Z.}~\bibnamefont {Preisler}},
  \bibinfo {author} {\bibfnamefont {F.}~\bibnamefont {Smallenburg}}, \bibinfo
  {author} {\bibfnamefont {M.}~\bibnamefont {Dijkstra}}, \ and\ \bibinfo
  {author} {\bibfnamefont {F.}~\bibnamefont {Sciortino}},\ }\href@noop {}
  {\bibfield  {journal} {\bibinfo  {journal} {J. Chem. Phys.}\ }\textbf
  {\bibinfo {volume} {138}},\ \bibinfo {eid} {164505} (\bibinfo {year}
  {2013})}\BibitemShut {NoStop}%
\bibitem [{\citenamefont {Beltran-Villegas}\ \emph {et~al.}(2014)\citenamefont
  {Beltran-Villegas}, \citenamefont {Schultz}, \citenamefont {Nguyen},
  \citenamefont {Glotzer},\ and\ \citenamefont {Larson}}]{janus_sedimentation}%
  \BibitemOpen
  \bibfield  {author} {\bibinfo {author} {\bibfnamefont {D.~J.}\ \bibnamefont
  {Beltran-Villegas}}, \bibinfo {author} {\bibfnamefont {B.~A.}\ \bibnamefont
  {Schultz}}, \bibinfo {author} {\bibfnamefont {N.~H.~P.}\ \bibnamefont
  {Nguyen}}, \bibinfo {author} {\bibfnamefont {S.~C.}\ \bibnamefont {Glotzer}},
  \ and\ \bibinfo {author} {\bibfnamefont {R.~G.}\ \bibnamefont {Larson}},\
  }\href {\doibase 10.1039/C3SM53136H} {\bibfield  {journal} {\bibinfo
  {journal} {Soft Matter}\ }\textbf {\bibinfo {volume} {10}},\ \bibinfo {pages}
  {4593} (\bibinfo {year} {2014})}\BibitemShut {NoStop}%
\bibitem [{\citenamefont {Russo}\ \emph {et~al.}(2011)\citenamefont {Russo},
  \citenamefont {Tavares}, \citenamefont {Teixeira}, \citenamefont {Telo~da
  Gama},\ and\ \citenamefont {Sciortino}}]{russo2011reentrant}%
  \BibitemOpen
  \bibfield  {author} {\bibinfo {author} {\bibfnamefont {J.}~\bibnamefont
  {Russo}}, \bibinfo {author} {\bibfnamefont {J.~M.}\ \bibnamefont {Tavares}},
  \bibinfo {author} {\bibfnamefont {P.~I.~C.}\ \bibnamefont {Teixeira}},
  \bibinfo {author} {\bibfnamefont {M.~M.}\ \bibnamefont {Telo~da Gama}}, \
  and\ \bibinfo {author} {\bibfnamefont {F.}~\bibnamefont {Sciortino}},\ }\href
  {\doibase 10.1103/PhysRevLett.106.085703} {\bibfield  {journal} {\bibinfo
  {journal} {Phys. Rev. Lett.}\ }\textbf {\bibinfo {volume} {106}},\ \bibinfo
  {pages} {085703} (\bibinfo {year} {2011})}\BibitemShut {NoStop}%
\bibitem [{\citenamefont {Rovigatti}\ \emph {et~al.}(2013)\citenamefont
  {Rovigatti}, \citenamefont {Tavares},\ and\ \citenamefont
  {Sciortino}}]{closed_loop}%
  \BibitemOpen
  \bibfield  {author} {\bibinfo {author} {\bibfnamefont {L.}~\bibnamefont
  {Rovigatti}}, \bibinfo {author} {\bibfnamefont {J.~M.}\ \bibnamefont
  {Tavares}}, \ and\ \bibinfo {author} {\bibfnamefont {F.}~\bibnamefont
  {Sciortino}},\ }\href {\doibase 10.1103/PhysRevLett.111.168302} {\bibfield
  {journal} {\bibinfo  {journal} {Phys. Rev. Lett.}\ }\textbf {\bibinfo
  {volume} {111}},\ \bibinfo {pages} {168302} (\bibinfo {year}
  {2013})}\BibitemShut {NoStop}%
\bibitem [{\citenamefont {Virnau}\ and\ \citenamefont
  {M\"{u}ller}(2004)}]{sus}%
  \BibitemOpen
  \bibfield  {author} {\bibinfo {author} {\bibfnamefont {P.}~\bibnamefont
  {Virnau}}\ and\ \bibinfo {author} {\bibfnamefont {M.}~\bibnamefont
  {M\"{u}ller}},\ }\href {\doibase 10.1063/1.1739216} {\bibfield  {journal}
  {\bibinfo  {journal} {J. Chem. Phys.}\ }\textbf {\bibinfo {volume} {120}},\
  \bibinfo {pages} {10925} (\bibinfo {year} {2004})}\BibitemShut {NoStop}%
\bibitem [{\citenamefont {Ferrenberg}\ and\ \citenamefont
  {Swendsen}(1989)}]{histogram_reweight}%
  \BibitemOpen
  \bibfield  {author} {\bibinfo {author} {\bibfnamefont {A.~M.}\ \bibnamefont
  {Ferrenberg}}\ and\ \bibinfo {author} {\bibfnamefont {R.~H.}\ \bibnamefont
  {Swendsen}},\ }\href {\doibase 10.1103/PhysRevLett.63.1195} {\bibfield
  {journal} {\bibinfo  {journal} {Phys. Rev. Lett.}\ }\textbf {\bibinfo
  {volume} {63}},\ \bibinfo {pages} {1195} (\bibinfo {year}
  {1989})}\BibitemShut {NoStop}%
\bibitem [{\citenamefont {Binder}\ \emph {et~al.}(2012)\citenamefont {Binder},
  \citenamefont {Block}, \citenamefont {Virnau},\ and\ \citenamefont
  {Tröster}}]{binder_spinodal}%
  \BibitemOpen
  \bibfield  {author} {\bibinfo {author} {\bibfnamefont {K.}~\bibnamefont
  {Binder}}, \bibinfo {author} {\bibfnamefont {B.~J.}\ \bibnamefont {Block}},
  \bibinfo {author} {\bibfnamefont {P.}~\bibnamefont {Virnau}}, \ and\ \bibinfo
  {author} {\bibfnamefont {A.}~\bibnamefont {Tröster}},\ }\href {\doibase
  http://dx.doi.org/10.1119/1.4754020} {\bibfield  {journal} {\bibinfo
  {journal} {Am. J. Phys.}\ }\textbf {\bibinfo {volume} {80}},\ \bibinfo
  {pages} {1099} (\bibinfo {year} {2012})}\BibitemShut {NoStop}%
\bibitem [{\citenamefont {Henderson}\ and\ \citenamefont
  {Speedy}(1987)}]{henderson1987temperature}%
  \BibitemOpen
  \bibfield  {author} {\bibinfo {author} {\bibfnamefont {S.~J.}\ \bibnamefont
  {Henderson}}\ and\ \bibinfo {author} {\bibfnamefont {R.~J.}\ \bibnamefont
  {Speedy}},\ }\href@noop {} {\bibfield  {journal} {\bibinfo  {journal} {J.
  Phys. Chem.}\ }\textbf {\bibinfo {volume} {91}},\ \bibinfo {pages} {3062}
  (\bibinfo {year} {1987})}\BibitemShut {NoStop}%
\bibitem [{\citenamefont {Wertheim}(1984)}]{Werth1}%
  \BibitemOpen
  \bibfield  {author} {\bibinfo {author} {\bibfnamefont {M.}~\bibnamefont
  {Wertheim}},\ }\href@noop {} {\bibfield  {journal} {\bibinfo  {journal} {J.
  Stat. Phys.}\ }\textbf {\bibinfo {volume} {35}},\ \bibinfo {pages} {19, ibid.
  35} (\bibinfo {year} {1984})}\BibitemShut {NoStop}%
\bibitem [{\citenamefont {Tavares}\ \emph {et~al.}(2014)\citenamefont
  {Tavares}, \citenamefont {Almarza},\ and\ \citenamefont {Telo~da
  Gama}}]{tavares}%
  \BibitemOpen
  \bibfield  {author} {\bibinfo {author} {\bibfnamefont {J.~M.}\ \bibnamefont
  {Tavares}}, \bibinfo {author} {\bibfnamefont {N.~G.}\ \bibnamefont
  {Almarza}}, \ and\ \bibinfo {author} {\bibfnamefont {M.~M.}\ \bibnamefont
  {Telo~da Gama}},\ }\href {\doibase http://dx.doi.org/10.1063/1.4863135}
  {\bibfield  {journal} {\bibinfo  {journal} {J. Chem. Phys.}\ }\textbf
  {\bibinfo {volume} {140}},\ \bibinfo {eid} {044905} (\bibinfo {year}
  {2014}),\ http://dx.doi.org/10.1063/1.4863135}\BibitemShut {NoStop}%
\bibitem [{\citenamefont {Zhang}\ \emph {et~al.}(2009)\citenamefont {Zhang},
  \citenamefont {Jin},\ and\ \citenamefont {Zhao}}]{Zhang2009}%
  \BibitemOpen
  \bibfield  {author} {\bibinfo {author} {\bibfnamefont {J.}~\bibnamefont
  {Zhang}}, \bibinfo {author} {\bibfnamefont {J.}~\bibnamefont {Jin}}, \ and\
  \bibinfo {author} {\bibfnamefont {H.}~\bibnamefont {Zhao}},\ }\href {\doibase
  10.1021/la9000279} {\bibfield  {journal} {\bibinfo  {journal} {Langmuir}\
  }\textbf {\bibinfo {volume} {25}},\ \bibinfo {pages} {6431} (\bibinfo {year}
  {2009})}\BibitemShut {NoStop}%
\bibitem [{\citenamefont {Andala}\ \emph {et~al.}(2012)\citenamefont {Andala},
  \citenamefont {Shin}, \citenamefont {Lee},\ and\ \citenamefont
  {Bishop}}]{Andala2012}%
  \BibitemOpen
  \bibfield  {author} {\bibinfo {author} {\bibfnamefont {D.~M.}\ \bibnamefont
  {Andala}}, \bibinfo {author} {\bibfnamefont {S.~H.~R.}\ \bibnamefont {Shin}},
  \bibinfo {author} {\bibfnamefont {H.-Y.}\ \bibnamefont {Lee}}, \ and\
  \bibinfo {author} {\bibfnamefont {K.~J.~M.}\ \bibnamefont {Bishop}},\ }\href
  {\doibase 10.1021/nn202556b} {\bibfield  {journal} {\bibinfo  {journal} {ACS
  Nano}\ }\textbf {\bibinfo {volume} {6}},\ \bibinfo {pages} {1044} (\bibinfo
  {year} {2012})}\BibitemShut {NoStop}%
\bibitem [{\citenamefont {Wang}\ \emph {et~al.}(2012)\citenamefont {Wang},
  \citenamefont {Wang}, \citenamefont {Breed}, \citenamefont {Manoharan},
  \citenamefont {Feng}, \citenamefont {Hollingsworth}, \citenamefont {Weck},\
  and\ \citenamefont {Pine}}]{Wang2012e}%
  \BibitemOpen
  \bibfield  {author} {\bibinfo {author} {\bibfnamefont {Y.}~\bibnamefont
  {Wang}}, \bibinfo {author} {\bibfnamefont {Y.}~\bibnamefont {Wang}}, \bibinfo
  {author} {\bibfnamefont {D.~R.}\ \bibnamefont {Breed}}, \bibinfo {author}
  {\bibfnamefont {V.~N.}\ \bibnamefont {Manoharan}}, \bibinfo {author}
  {\bibfnamefont {L.}~\bibnamefont {Feng}}, \bibinfo {author} {\bibfnamefont
  {A.~D.}\ \bibnamefont {Hollingsworth}}, \bibinfo {author} {\bibfnamefont
  {M.}~\bibnamefont {Weck}}, \ and\ \bibinfo {author} {\bibfnamefont {D.~J.}\
  \bibnamefont {Pine}},\ }\href {\doibase 10.1038/nature11564} {\bibfield
  {journal} {\bibinfo  {journal} {Nature}\ }\textbf {\bibinfo {volume} {490}},\
  \bibinfo {pages} {51} (\bibinfo {year} {2012})}\BibitemShut {NoStop}%
\bibitem [{\citenamefont {Yi}\ \emph {et~al.}(2013)\citenamefont {Yi},
  \citenamefont {Pine},\ and\ \citenamefont {Sacanna}}]{sacanna_review}%
  \BibitemOpen
  \bibfield  {author} {\bibinfo {author} {\bibfnamefont {G.-R.}\ \bibnamefont
  {Yi}}, \bibinfo {author} {\bibfnamefont {D.~J.}\ \bibnamefont {Pine}}, \ and\
  \bibinfo {author} {\bibfnamefont {S.}~\bibnamefont {Sacanna}},\ }\href@noop
  {} {\bibfield  {journal} {\bibinfo  {journal} {J. Phys.: Condens. Matter}\
  }\textbf {\bibinfo {volume} {25}},\ \bibinfo {pages} {193101} (\bibinfo
  {year} {2013})}\BibitemShut {NoStop}%
\bibitem [{\citenamefont {Hansen}\ and\ \citenamefont
  {McDonald}(1990)}]{hansen1990theory}%
  \BibitemOpen
  \bibfield  {author} {\bibinfo {author} {\bibfnamefont {J.-P.}\ \bibnamefont
  {Hansen}}\ and\ \bibinfo {author} {\bibfnamefont {I.~R.}\ \bibnamefont
  {McDonald}},\ }\href@noop {} {\emph {\bibinfo {title} {Theory of simple
  liquids}}}\ (\bibinfo  {publisher} {Elsevier},\ \bibinfo {year}
  {1990})\BibitemShut {NoStop}%
\end{thebibliography}%

\section*{Supplementary Information}

\section{Models and methods}
In the Kern-Frankel~\cite{Kern2003} model, 
given two particles $i$ and $j$ separated by a distance vector $\mathbf{r}_{ij}$, the interaction between two patches $\alpha$ and $\beta$ on their surface is given by
\begin{equation}
u_{\rm pp}(\mathbf{r}_{ij}, \mathbf{\hat{n}}_\alpha, \mathbf{\hat{n}}_\beta) = u_{\rm SW}(r_{ij}) f(\mathbf{r}_{ij}, \mathbf{\hat{n}}_\alpha, \mathbf{\hat{n}}_\beta)
\end{equation}
where $r_{ij} = |\mathbf{r}_{ij}|$, $u_{\rm SW}$ is an isotropic square well term of range $\sigma + \delta_{\alpha\beta}$ and depth $\epsilon_{\alpha\beta}$, $\mathbf{\hat{n}}_\kappa$ is the unit vector connecting patch $\kappa$ to the centre of its particle and $f$ is an orientation-dependent term which takes the form
\begin{equation}
f(\mathbf{r}_{ij}, \mathbf{\hat{n}}_\alpha, \mathbf{\hat{n}}_\beta) = \left\{ \begin{array}{rl}  
1 & \mathrm{if} \left\{ \begin{array}{rl} 
\mathbf{\hat{r}}_{ij} \cdot \mathbf{\hat{n}}_\alpha > \cos{\theta^\mathrm{max}_{\alpha\beta}}\\
\mathbf{\hat{r}}_{ji} \cdot \mathbf{\hat{n}}_\beta > \cos{\theta^\mathrm{max}_{\alpha\beta}}
\end{array}\right.\\
0 & \mathrm{otherwise}
\end{array} \right.
\end{equation}
where $\cos{\theta^\mathrm{max}_{\alpha\beta}}$ controls the angular width of the patch-patch interaction.

The KF parameters of the Janus model are $\epsilon_{\alpha\beta} = \epsilon = 1$, $\delta_{\alpha\beta} = \delta = 0.5$ and $\cos{\theta^\mathrm{max}_{\alpha\beta}} = \cos{\theta^\mathrm{max}} = 0$. We simulate a box of linear size $L = 20 \sigma$.

The KF parameters for the 2A9B model are $\epsilon_{BB} = 0$, $\epsilon_{AA} = 1$, $\epsilon_{AB} = 0.37$, $\delta_{AA} = 0.01$, $\delta_{AB} = 0.545$, $\cos{\theta^\mathrm{max}_{AA}} = 0.94549525$, $\cos{\theta^\mathrm{max}_{AB}} = 0.9461856$. We note that these values enforce the single-bond-per-patch condition for all bond types. We simulate a box of linear size $L = 14 \sigma$.

The liquid branch of the equation of state is splined under tension to increase accuracy. The density maxima are found by considering all the intersections between two $P(\rho,T)$ evaluated at different but nearby temperatures, $T_1$ and $T_2$. For each intersection the pressure of the maximum is considered to be the pressure at which the curves cross, whereas the temperature of the maximum is taken as $T_M \approx (T_1 + T_2) / 2$. 

\section{Evaluating the pressure in the grand-canonical ensemble}
\label{app:P}

In the grand-canonical ensemble 
the compressibility $K_T$ is positive-defined, being $\rho k_B T K_T = \langle\Delta N^2\rangle/\langle N \rangle$, where $\rho = N/V$ and the right-hand side is the relative root-mean-squared deviation of the probability  ${\cal N}(N)$ of finding $N$ particles in the simulated volume $V$. ${\cal N}(N)$ is linked to both the grand-canonical ($\mathcal{Z}$) and canonical ($\mathcal{Q}_N$) partition functions \textit{via}

\begin{equation}
{\cal N}(N) = \frac{z^N \mathcal{Q}_N(V, T)}{\mathcal{Z}(\mu, V, T)},
\end{equation}
which implies that ${\cal N}(0) = \mathcal{Z}^{-1}(\mu, V, T)$ and thus that $\mathcal{Q}_N(V, T) = {\cal N}(N) / (z^N {\cal N}(0))$. The latter relation allows us to compute quantities in the canonical ensemble.
We will refer to these quantities using the superscript $^c$. For example,  the pressure can be computed via   the relation that links the Gibbs $G$ and Helmholtz $F$ free energies:

\begin{equation}
\label{eq:G}
\beta G = \beta \mu^c N = \beta F + \beta P^cV = -\log(\mathcal{Q}_{N}) + \beta P^cV.
\end{equation}

 $\mu^c$ can be evaluated using Widom's formula~\cite{hansen1990theory}

\begin{equation}
\label{eq:mu}
\beta \mu^c \simeq -\log\left(\frac{\mathcal{Q}_{N+1}}{\mathcal{Q}_{N}}\right) = -\log\left(\frac{{\cal N}(N+1)}{z {\cal N}(N)}\right)
\end{equation}

If we substitute Eq.(\ref{eq:mu}) in Eq.(\ref{eq:G}) and resolve for $\beta P^c$ we obtain

\begin{equation}
\beta P^c = \frac{- N \log\left(\frac{\mathcal{Q}_{N+1}}{\mathcal{Q}_{N}}\right) + \log(\mathcal{Q}_{N})}{V}
\end{equation}

which can be expressed as a function of ${\cal N}(N)$ as

\begin{equation}
\beta P^c = \frac{- N \log\left( \frac{{\cal N}(N+1)}{{\cal N}(N)} \right) + \log\left( \frac{{\cal N}(N)}{{\cal N}(0)} \right)}{V}.
\label{pressure}
\end{equation}

We note that the last equation does not depend on the chemical potential $\mu$ used for the grand-canonical simulation. Indeed, changing the chemical potential from $\mu$ to $\mu'$ would imply a change in ${\cal N}(N)$ given by

\begin{equation}
 p_\mu(N) \longrightarrow p_{\mu'}(N) = p_\mu(N) e^{\beta\Delta\mu N}
\end{equation}
where $\Delta\mu=\mu'-\mu$ and the labels $\mu$, $\mu'$ refer to the probability distribution calculated with grand-canonical simulation performed with chemical potential $\mu$ and $\mu'$ respectively. 
Replacing $p_{\mu'}(N)$ in Eq. (\ref{pressure}) does not affect $P^c$.

\end{document}